\documentclass{aa}  

\usepackage{graphicx}
\usepackage{txfonts}
\def\gtsima{$\; \buildrel > \over \sim \;$}
\def\ltsima{$\; \buildrel < \over \sim \;$}
\def\prosima{$\; \buildrel \propto \over \sim \;$}
\def\gsim{\lower.5ex\hbox{\gtsima}}
\def\lsim{\lower.5ex\hbox{\ltsima}}
\def\simgt{\lower.5ex\hbox{\gtsima}}
\def\simlt{\lower.5ex\hbox{\ltsima}}
\def\simpr{\lower.5ex\hbox{\prosima}}

\def\h1{$h^{-1}$}
\def\eeq{\end{equation}}
\def\beq{\begin{equation}}

\begin{document}

 \title{CO excitation of normal star forming galaxies out to $z=1.5$ as regulated by the properties
 of their interstellar medium}


   \author{E. Daddi\inst{1}
        \and H. Dannerbauer\inst{2}
   	\and D. Liu\inst{1,3,4}
	\and M. Aravena\inst{5}
	\and F. Bournaud\inst{1}
 	\and F. Walter\inst{6}
        \and D. Riechers\inst{7} 
        \and G. Magdis\inst{8,15}
        \and M. Sargent\inst{9}
        \and M. B\'{e}thermin\inst{10}
        \and C. Carilli\inst{11}
        \and A. Cibinel\inst{9}
        \and M. Dickinson\inst{12}
        \and D. Elbaz\inst{1}
        \and Y. Gao\inst{3}
        \and R. Gobat\inst{1,13}
        \and J. Hodge\inst{11}
        \and M. Krips\inst{14}
          }
\institute{CEA Saclay, Laboratoire AIM-CNRS-Universit\`e Paris Diderot, Irfu/SAp,
Orme des Merisiers, F-91191 Gif-sur-Yvette, France
 \email{edaddi@cea.fr}
        \and Universit\"{a}t Wien, Institut f\"{u}r Astrophysik, T\"{u}rkenschanzstrasse 17, 1180 Vienna, Austria
        \and Purple Mountain Observatory \& Key Laboratory for Radio Astronomy, Chinese Academy of Sciences, Nanjing 210008, China
        \and Graduate University of the Chinese Academy of Sciences, 19A Yuquan Road, Shijingshan District, Beijing 10049, China
        \and N\'ucleo de Astronom\'{\i}a, Facultad de Ingenier\'{\i}a, Universidad Diego Portales, Av. Ej\'ercito Libertador 441, Santiago Chile
        \and Max-Planck Institute for Astronomy, K\"{o}nigstuhl 17, 69117 Heidelberg, Germany
                \and Department of Astronomy, Cornell University, 220 Space Sciences Building, Ithaca, NY 14853, USA
        \and Department of Physics, University of Oxford, Keble Road, Oxford OX1 3RH, UK
        \and Astronomy Center, Department of Physics and Astronomy, University of Sussex, Brighton BN1 9QH, UK
        \and European Southern Observatory, Karl-Schwarzschild-Str. 2, 85748 Garching, Germany
        \and National Radio Astronomy Observatory, P.O. Box O, Socorro, NM 87801, USA
        \and National Optical Astronomical Observatory, 950 North Cherry Avenue, Tucson, AZ 85719, USA
        \and School of Physics, Korea Institute for Advanced Study, Heogiro 85, Seoul 130-722, Republic of Korea
       \and IRAM - Institut de RadioAstronomie Millimetrique, 300 rue de la Piscine, 38406 Saint Martin d'Heres, France
       \and Institute for Astronomy, Astrophysics, Space Applications and Remote Sensing, National Observatory of Athens, GR-15236 Athens, Greece
        }
     
     \titlerunning{CO excitation in $z=1.5$ disk galaxies}
     \authorrunning{E. Daddi et al.}        
     

   \date{Received ; accepted }

 
  \abstract
 {
 We investigate the CO excitation  of  normal (near-IR selected BzK) star forming (SF) disk galaxies at $z=1.5$ using IRAM Plateau de Bure  observations of the
  CO[2-1], CO[3-2] and CO[5-4] transitions for 4 galaxies,  including VLA observations of CO[1-0] for 3 of them, with the aim of constraining the average state of H$_2$ gas. Exploiting prior knowledge of the velocity range, spatial extent and size of the CO emission we measure reliable line fluxes with S/N$>4$--7  for individual transitions.   While the average CO Spectral Line Energy Distribution 
  (SLED)  has a sub-thermal excitation similar to the Milky Way (MW) up to CO[3-2], we show that the average CO[5-4] emission is four times stronger than assuming MW excitation. This demonstrates the presence of an additional component of more excited, denser and possibly warmer molecular gas.   The ratio of CO[5-4] to lower-J CO emission is however lower than in local (U)LIRGs and high-redshift starbursting SMGs, and appears to correlate closely with the average intensity 
  of the radiation field $<U>$ and with the star formation surface density, but not with the SF efficiency (SFE). 
  The luminosity of the CO[5-4] transition is found to correlate linearly with the bolometric infrared luminosity over 4 orders of magnitudes. For this transition, $z=1.5$ BzK galaxies follow the same linear trend
 as local spirals and (U)LIRGs and high redshift star bursting sub-millimeter galaxies. The  CO[5-4] luminosity is thus empirically
  related to the dense  gas, and might be a more convenient way to probe it than standard high--density tracers that are much fainter than CO. 
  We see excitation variations among our sample galaxies, that can be linked  to their evolutionary state and clumpiness in optical rest frame images. In one galaxy we see spatially resolved excitation variations, where the more highly excited part of the galaxy corresponds to the location of massive SF clumps. This provides support to models
  that suggest that giant clumps are the main source of the high excitation CO emission in high redshift disk-like galaxies.
   }


   \maketitle
%

\section{Introduction}

The nature of star formation in high redshift galaxies and the physical properties of the interstellar medium (ISM) in which it is taking place are not yet very well understood. Empirically, we now know that most star forming galaxies have higher specific SFRs (sSFR) when going to higher redshifts, all the way to redshift 3 and possibly beyond (see. e.g., Sargent et al 2014, Madau \& Dickinson 2014, for up to date compilations). This rise of sSFR appears to take place in a rather well-ordered way.
The increase is roughly independent of stellar mass at least for massive galaxies in the range $9.5<{\rm log} M/M_\odot<11.5$. This is the so-called 'Main Sequence' (MS) paradigm in which star forming galaxies define a tight correlation between their stellar mass and SFR with typically a factor of two dispersion (Noeske et al 2007; Elbaz et al 2007; 2011; Daddi et al 2007; 2009a; Pannella et al 2009; Karim et al 2011; Whitaker et al 2012; Magdis et al. 2010; Rodighiero et al 2011; 2014; Schreiber et al. 2014). When studying the most luminous systems (e.g., selected by means of their ultra-high infrared bolometric luminosity -- $L_{\rm IR}(8-1000\mu$m), beyond $10^{12}L_\odot$ at $z=0$ or beyond $10^{13}L_\odot$ at $z=2$; Sargent et al 2012) one is bound to find a rarer kind of galaxy with more extreme sSFR likely to be in a short lived,  possibly merging driven, starburst phase. Such objects are generally outliers to the MS (where, instead,  disk like systems typically live; Wuyts et al 2011; Salmi et al 2012; Forster-Schreiber et al 2009) and account for a modest fraction of the global SFR density (e.g., 10--20\%; Rodighiero et al 2011; Sargent et al 2012; Hopkins et al 2011). It is important  to distinguish conceptually between both kinds of galaxies (i.e. between MS \& SB galaxies) in order to achieve a global understanding of star forming galaxies through cosmic time.

When concentrating on MS galaxies  (or, almost equivalently, studying near-IR or optically selected galaxy samples) we are starting now to get a fairly detailed empirical picture of the redshift evolution of some of their key physical properties, which change in parallel to the rise of their sSFRs. First, their gas fractions appear to be rising substantially at least from $z=0$ to 2, going from some roughly 5-10\% at $z=0$ to 40-50\% at $z=2$ for stellar masses around a few $10^{10}M_\odot$.
This result has been obtained both from investigation of CO emission lines largely from the IRAM Plateau de Bure interferometer 
(Daddi et al 2008; 2010a; D10 hereafter;   Tacconi et al 2010; 2013; Geach et al 2011) and confirmed by gas mass estimates obtained via the conversion of dust masses  (Magdis et al 2011; 2012; Santini et al 2014; Scoville et al 2014; Magnelli et al 2013; Genzel et al 2014). Their ISM properties are also changing: most notably their dust temperatures are getting warmer as reflected by the increase with redshift of the average intensity of the radiation field $<U>$ (Magdis et al 2012; Magnelli et al 2013; B\'{e}thermin et al 2014; Genzel et al 2014). Their metal content is decreasing, with metallicity at fixed stellar mass declining  by more
than a factor of 2 towards $z=2$ (Erb et al 2006; Maier et al 2014; Steidel et al 2014), and with some hints existing of a much more rapid decrease at higher redshifts (Troncoso et al 2014). On the other hand, the ISM dust attenuation properties are not changing much with only a modest rise of the attenuation at fixed stellar mass (e.g., Pannella et al 2014; Burgarella et al 2013) that is justified by the increase of the dust to stellar mass ratio (Tan et al 2014). However, the difference between the attenuation of the HII regions and the stellar continuum appears to be reduced with respect to the local Universe (Kashino et al 2013; Price et al 2013; Pannella et al 2014). Finally, MS galaxies from $z=0$ to beyond 2 appear to define a non-linear trend in the SFR-M$_{\rm gas}$ plane (either for integrated or specific quantities in the Schmidt-Kennicutt relation; KS; Daddi et al 2010b; Genzel et al 2010; Sargent et al 2014), so that high redshift galaxies, which have higher SFR, also have higher star formation efficiencies (SFE $=SFR/M_{\rm gas}$). 
Yet, this wealth of empirical information do not still suffice to clarify the physical properties of the ISM. Studies of emission line ratios suggest that the properties of the  ionizing radiation field are different at $z=2$, with harder and possibly more intense spectra (Steidel et al 2014) that are in qualitative agreement 
with the observations of higher $<U>$ (dust is a bolometric tracer and thus cannot distinguish harder vs more intense fields). This might affect the kinetic temperature of the gas. At the same time the gas densities might be changing, given the higher gas fractions and possibly decreasing galaxy sizes towards 
higher redshifts (e.g., van der Wel et al. 2014). 

One important piece of information to gain further insight into the ISM properties is the study of the excitation of the CO emission, i.e. the relative luminosity ratio of 
CO lines with different rotational quantum number ($J_{\rm upper}$). CO is the most luminous tracer of molecular hydrogen (H$_2$), the fuel for star formation
in galaxies. The specific distribution of intensity of high-J versus low-J CO is sensitive to the H$_2$ gas density and temperature, albeit in a somewhat degenerate way, providing insights into the physical properties of the gas. The effective critical densities for excitation rise from $\sim300$~cm$^{-3}$ for CO[1-0], which is thus most sensitive to the total gas reservoir including more diffuse components, to e.g. $\sim10^4$ ~cm$^{-3}$ for CO[5-4], so that $J>1$ transitions are more and more sensitive to the denser star forming gas (e.g., Solomon \& van den Bout 2005).
On the other hand, the study of the CO excitation properties of high redshift MS galaxies is interesting also empirically as a means to further understand their nature and the possible role of mergers/starbursts in driving their properties. In fact  it is well known locally that, empirically/observationally,  different CO excitation properties characterize spiral galaxies as opposed to merging driven (U)LIRGs. The latter are much more highly excited in their high-J CO transitions (Weiss et al 2007; Papadopoulos et al 2012; P12 hereafter).
However,  CO  SLEDs  by
  themselves are not necessarily good indicators  of the nature of star formation and galaxy types. No matter if
  they are vigorously SF disks, or mergers, CO SLEDs of warm gas will
  be similar.  It is primarily  the dense to total  gas fraction  that  can be a  SF mode indicator  (merger vs
  disk; e.g., Daddi et al 2010b; Zhang et al. 2014), if it can  be constrained by
  the available  SLEDs.
 Not much is known observationally about the CO excitation properties of indisputably normal MS galaxies high redshift. The most detailed information is available for IR-luminous systems selected either as SMGs (e.g., Bothwell et al 2013; Riechers et al 2013) or strong lenses (Danielson et al 2011;  Cox et al 2011; Combes et al 2012; Scott et al 2011; Spilker et al 2014). In a pilot study Dannerbauer et al (2009) presented CO[3-2] observations of one $z=1.5$ galaxy (BzK-21000, further discussed in this paper) previously detected in CO[2-1], finding a low CO[3-2]/CO[2-1] ratio fairly consistent with the Milky~Way (MW). Aravena et al (2014) presents 
CO[1-0] Karl G. Jansky Very Large Array (VLA) observations of 3 BzK-selected  and one UV-selected galaxies which, compared to CO[2-1] or CO[3-2] observations of the same objects, display fairly low excitation ratios. Not much more is known about the excitation properties of MS-selected galaxies for higher J transitions, i.e. those most sensitive to warm/dense gas, a situation we aim to start remedying with the present study with new CO[5-4] observations. This is particularly important as the study of SLEDs up to CO[3-2] only can be fairly degenerate as the type of underlying ISM conditions (e.g., Dannerbauer et al 2009; Papadopoulos et al 2014).

There are empirical reasons to expect evolution in the CO spectral energy distributions (SLEDs) of MS disk-like galaxies at high redshift compared to local spirals and the MW, mostly because of the already discussed evolution in related physical properties. For example, the rise in SFE suggests the parallel increase of the fraction of dense versus total molecular gas  (e.g., Daddi et al. 2010b). The increase of the mean radiation field intensity $<U>$ is plausibly directly reflected in rising gas temperatures. All of this would suggest the idea that CO SLEDs might not be identical to the MW SLEDs
when going to higher J-transition, with some higher level of gas excitation likely being present. This kind of qualitative assessment agrees with theoretical predictions. 
For example, P12 predict  stronger high-J emission heated by turbulence and/or cosmic rays in these systems with higher SFEs than local spirals.
Similarly, Narayanan \& Krumholz (2014) suggest that the CO SLED is driven by the SFR surface density ($\Sigma_{\rm SFR}$), with galaxies with highest $\Sigma_{\rm SFR}$ having more excited high-J components. More recently, fully detailed very high resolution numerical hydrodynamical simulations of high-z disks have been used by Bournaud et al (2015; B15 hereafter) to predict the evolution of CO excitation, confirming the expectation for a significant high excitation component beyond CO[3-2]. { Semi-analytic models also predict increasing  CO excitation towards $z\sim2$ (Popping et al. 2014; Lagos et al 2012).}

In this paper we present CO[5-4] and CO[3-2] observations of 4 BzK galaxies at $z=1.5$. Coupled with existing measurements at CO[2-1] and CO[1-0] they allow us to derive  CO SLEDs for these objects extending  to CO[5-4] and to study their excitation properties,  and compare these with other physical parameters and with model predictions. The paper is organized as follows: in Section~\ref{obs}~and~\ref{mes} we present the dataset used and the measurements performed on our data. In Section \ref{sleds} we discuss results on the CO SLEDs
of the BzK galaxies, in comparison to other galaxy populations at low and high redshift and to theoretical models. Section~\ref{lvg} discusses
the physical constraints on the gas density and temperature that can be obtained from our data using Large Velocity Gradient (LVG) modelling. We discuss our findings in Section~\ref{disc}, where we try to identify the main driver of CO excitation and find that the mean radiation field intensity $<U>$ appears to be the best tracer together with $\Sigma_{\rm SFR}$. We also find that the BzK galaxies obey a linear correlation between the CO[5-4] luminosity and $L_{\rm IR}$ which is also followed by other galaxy populations. Finally,  we discuss the evidence that the CO excitation is linked to the presence of giant clumps. 
In the paper we use a Chabrier (2003) IMF when specifying SFRs and stellar masses,  and adopt a standard WMAP cosmology.

\section{Observations}\label{obs}

The targets of our observations are  four near-IR-selected galaxies with high S/N detections of CO[2-1] taken from D10: BzK-4171, BzK-16000, BzK-17999 and BzK-21000, 
with redshifts spanning the range $1.414<z<1.523$ and SFRs placing them in the star forming MS. There is no sign of AGN emission in these galaxies, from the 2~Ms Chandra data, radio 1.4~GHz continuum, optical spectroscopy, Spitzer IRAC colours, IRS Spitzer spectroscopy in 4171 and 21000, and mid- to far-IR decomposition (D07, Pope et al 2013; Daddi et al in preparation). 
These galaxies had been observed in CO[2-1] in at least two different configurations with the IRAM PdBI, 
including an extended configuration (B and/or A), that revealed resolved SF activity with the resulting data allowing  to constraint their CO[2-1] spatial sizes (D10)\footnote{The sample of D10 contains two more BzK sources that were observed only in one configuration and had thus lower S/N CO[2-1] detections.}. The fairly accurate CO[2-1] fluxes already available for these galaxies were an excellent starting point to investigate their CO excitation properties with the follow-up of different CO transitions.

\subsection{IRAM PdBI}

We observed our targets in the  CO[3-2] (345.796~GHz rest frame, redshifted into the 2mm band) and CO[5-4] (576.267~GHz rest frame, redshifted into the 1mm band) transitions during several observing campaigns of the IRAM PdBI, spanning the Winter~2008 to Winter~2011 IRAM semesters (see~Table~\ref{tab:1}). All observations were obtained in compact configurations given that we were 
mainly interested in the total fluxes for this detection experiment, resulting in synthesized beams of 4--5$''$ for CO[3-2] and 2--3$''$ for CO[5-4]
(Table~\ref{tab:1}). 
All sources are expected to remain unresolved in the data at this resolution. The CO[3-2] observations of BzK-21000 were the first to be obtained and have already been published in Dannerbauer et al (2009),
and are also used in this paper. The CO[3-2] observations of all four sources and the CO[5-4] data of BzK-21000 were obtained with the previous generation IRAM correlator, with a total bandwidth of 1~GHz,  while for the other three galaxies  the CO[5-4] observations performed since 2011  benefited from the new Widex correlator with a total bandwidth of 3.6~GHz.  All data were taken in  dual polarization mode. The combined on-source time is 82~hours of 6-antenna equivalent time (part of the data was taken with 5 antennas only), corresponding to a total of about 130~hours of PdBI observing time when considering overheads.
The data were reduced and calibrated using the GILDAS software package CLIC. The final data, imaged
using natural weighting, has typical noise in the range of 250--450 $\mu$Jy integrated
over 400~km~s$^{-1}$ bandwidth, representative of the line-widths of the target BzK galaxies. This corresponds to roughly 0.15~Jy~km~s$^{-1}$ when integrated over the velocity range. 
For two of the sources we slightly offsetted the pointing of the CO[3-2] observations in order to gather information also on interesting nearby sources  (GN10 for BzK-16000; Daddi 
et al 2009b; and the GN20/GN20.2ab sources close to BzK-21000; Daddi  et al 2009a; Carilli et al 2010; Hodge et al 2012; 2013; 2015; Tan et al 2014).  The sensitivity loss due to primary beam attenuation at off-center position was minimal, only 10\% at the position of the BzK galaxies. We correct all measurements for the flux loss. 
We expect that fluxes are affected by systematic calibration uncertainties at less than a level of $\pm10$\%, based on well studied and frequently observed calibrators (MWC349 primarily) and on the monitored efficiency of PdBI antennas. This is  below our  measurement errors.

\begin{table*}
{\footnotesize
\caption{Observation summary table}
\label{tab:1}
\begin{tabular}{l c c c c c c c c c}
\hline\hline
Source & RA$^1$ & DEC$^1$ & Transition & Frequency & Obs.dates & T$_{int}$$^2$ & Combined beam$^3$ & rms \\

 & J2000 & J2000 &  & GHz&  & hr. & &  $\mu$Jy$^4$\\
\hline
BzK-4171 & 12:36:26.520 & 62:08:35.48   & CO[3--2] & 140.282 & 04/2009 & 12.3 & 5.0$''\times 3.2''$, $PA=52^o$ & 290 \\
          & &  & CO[5--4]& 233.780 & 01-09/2011  & 11.5 & 2.8$''\times 2.0''$, $PA=113^o$ & 460 \\
\hline
BzK-16000 & 12:36:30.114 & 62:14:28.29   & CO[3--2] &137.057 & 04/2009  & 9.8 & $5.2'' \times 3.5'', PA=48^o$ & 410 \\
           & & & CO[5--4]& 228.225 & 04-07/2011  & 6.8 & $2.9'' \times 2.2'', PA=80^o$ & 390 \\
\hline
BzK-17999 & 12:37:51.827 & 62:15:19.90   & CO[3--2]& 143.246  & 04-05/2009   & 10.7 & $4.3'' \times 3.4'', PA=75^o$ & 370  \\
          & &  & CO[5--4] & 238.719  & 05-10/2011 & 12.9 & $2.5'' \times 2.1'', PA=92^o$ & 310 \\
\hline
BzK-21000 & 12:37:10.642 & 62:22:34.35  & CO[3--2]$^{5}$ & 136.970 & 05-06/2008  & 9.7 &  $4.0'' \times 3.1'', PA=72^o$ & 260  \\

& & & CO[5--4] & 228.574 & 05-10/2009  & 8.7 &  $2.9'' \times 2.0'', PA=50^o$ & 490  \\
\hline
\end{tabular}\\
Notes:\\
1: The coordinates correspond to the fixed positions for our CO extractions, as detailed in section \ref{sec:meas1}. \\
2: Equivalent six-antennas on source integration time. \\
3: Beam resulting from combination of all available configuration and imaging with natural weighting. These are the
beams sizes displayed in Figs.~\ref{fig:co32}~and~\ref{fig:co54}.\\
4: Noise per beam averaged over a spectral range corresponding to 400~km~s$^{-1}$.\\
5: These data have already been presented in Dannerbauer et al.~(2009).
}
\end{table*}

   \begin{figure*}
   \centering
   \includegraphics[width=18.5cm]{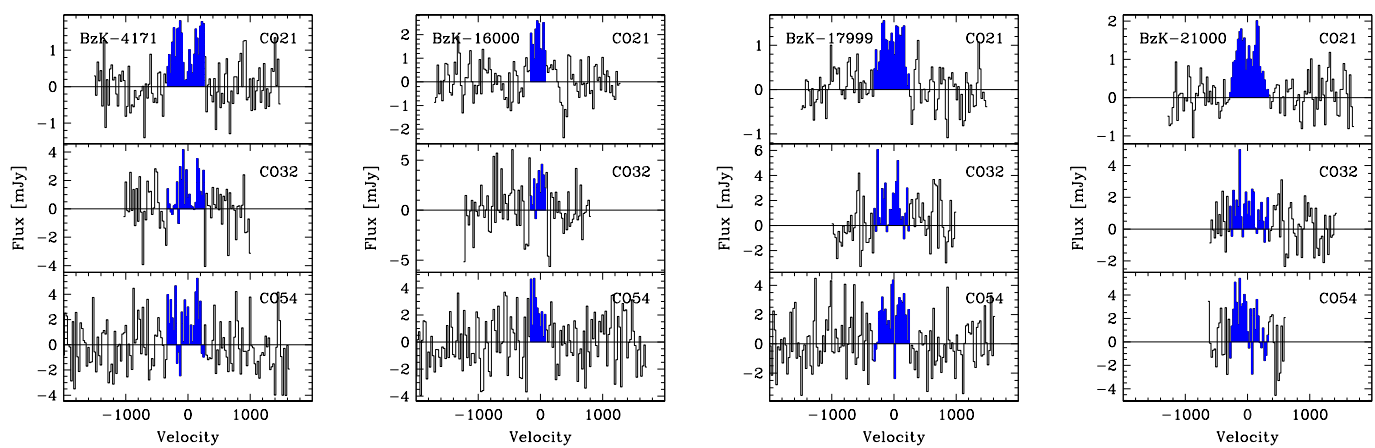}
   \caption{For each galaxy in our sample we show multi-transition spectra of CO[2-1] (top), CO[3-2] (middle) and CO[5-4] (bottom), extracted fitting circular Gaussian models convolved with the beam at the position of the galaxies, in step of 25~km~s$^{-1}$. The CO[2-1] data is from D10, the CO[3-2]
   data for BzK-21000 is from Dannerbauer et al. (2009).
   The shaded regions 
   correspond to the velocity range (kept fixed for all transitions of a given galaxy), over which fluxes are measured. Continuum levels as measured in the data
   and constrained from our multi-wavelength modelling have been subtracted. Zero velocity corresponds to the accurate CO[2-1] redshifts from D10.
			   }
              \label{fig:do3}%
    \end{figure*}
    
       \begin{figure*}
   \centering
   \includegraphics[width=18.5cm]{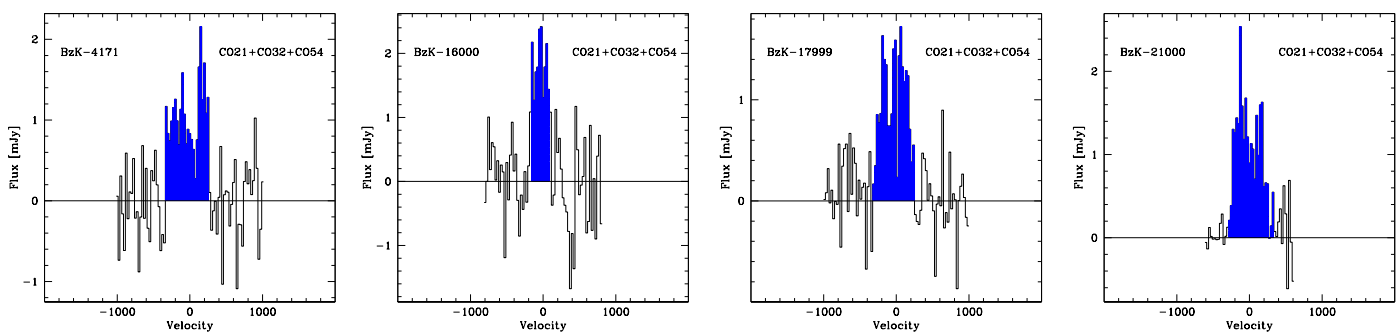}
   \caption{CO spectra averaged over the CO[2-1], CO[3-2], CO[5-4] transitions for each galaxy, after continuum subtraction, weighting with the S/N in each transition.
			   }
              \label{fig:doWei}%
    \end{figure*}

\subsection{VLA}

Three of our sources were observed in their CO[1-0] emission (115.271~GHz rest frame, redshifted to $\sim6.6$mm) at the VLA: BzK-4171, BzK-16000, and BzK-21000 (Aravena et al 2010; 2014). No CO[1-0] data is available for BzK-17999. A first observational campaign was conducted with the VLA during 2009 in the C and D configurations using 2 50~MHz channels, with results published by Aravena et al  (2010). The same 3 targets were re-observed with the VLA during 2010--2011 using the WIDAR correlator covering the expected line frequencies with 128 2~MHz channels. The new VLA observations are described in detail
in Hodge et al. (2012), and measurements for the BzK galaxies were published in Aravena et al (2014). In this paper we use the combination of new and old VLA observations 
to extend the CO SLEDs to the CO[1-0] transitions, performing new flux measurements for consistency with the IRAM data.   Resulting spatial resolutions were of the order of 1.5--2$''$ (Aravena et al 2010; 2014), except for BzK-21000 which benefited from part of the data being taken at a higher resolution. The VLA images of our targets are shown in Aravena et al (2010; 2014).

   \begin{figure}
   \centering
   \includegraphics[width=9cm]{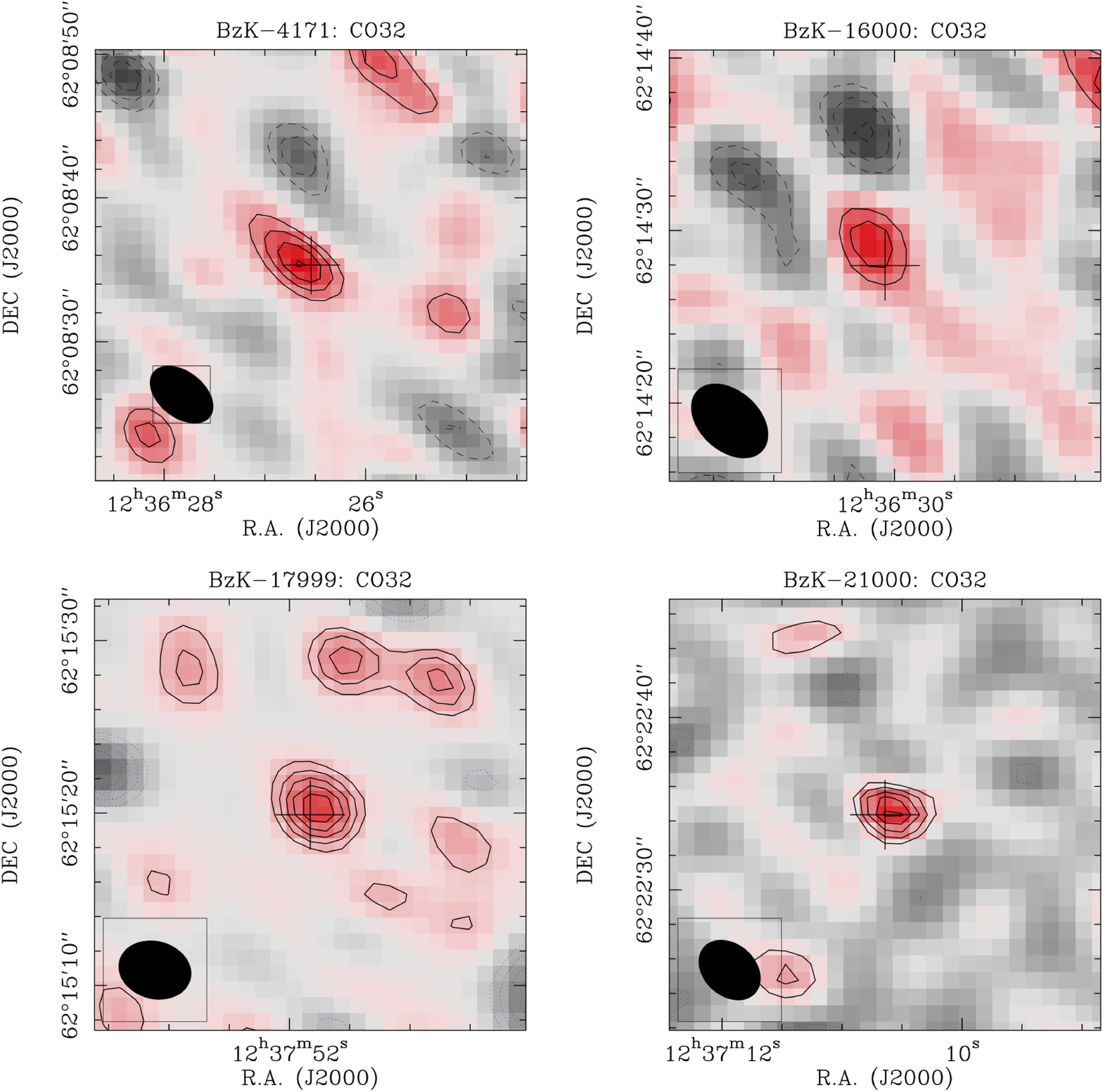}
   \caption{Velocity averaged maps of the CO[3-2] emission of our galaxies, using natural weighting. Contours levels start at $\pm2\sigma$ and increase in step of $1\sigma$. The size and orientation of the beam is indicated in the bottom-left corner. The cross in each panel is centered at the  CO positions from Table~\ref{tab:1}.
			   }
              \label{fig:co32}%
    \end{figure}
   \begin{figure}
   \centering
   \includegraphics[width=9cm]{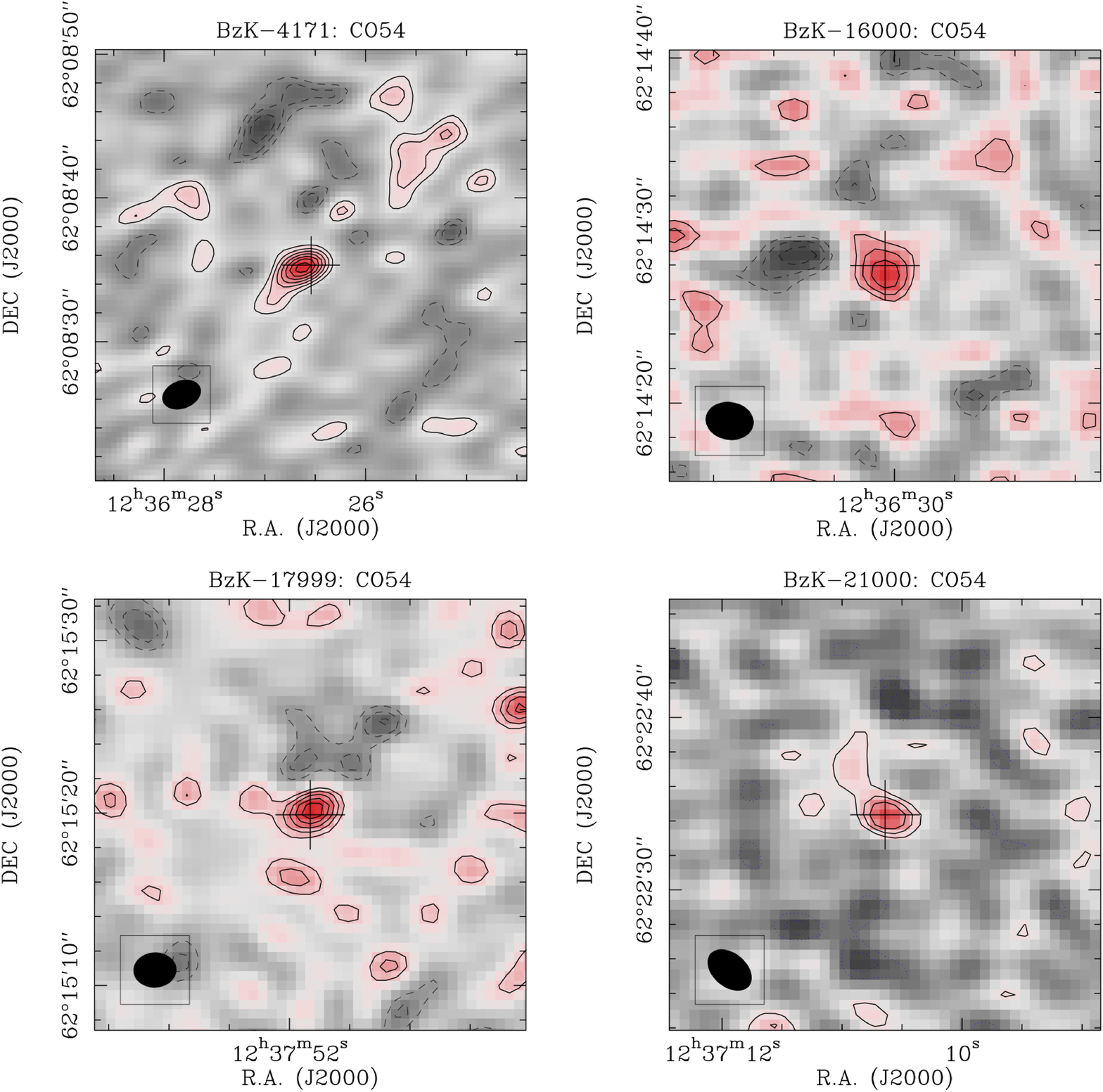}
   \caption{The same of Fig.~\ref{fig:co32} but for the CO[5-4] observations. Continuum emission has been subtracted here, as discussed
   in the text.
			   }
              \label{fig:co54}%
    \end{figure}

\section{Measurements}\label{mes}

Interferometric data are effectively 3D cubes including spatial and velocity information. Measuring a CO flux  involves integrating the data over a certain velocity range and afterwards extracting the flux with assumptions on the spatial profile. Hence, the global error connected to absolute flux measurements can be substantially affected by the possible ambiguity in choosing the velocity range and spatial position and extent of the source. In order to build CO SLEDs 
as accurately as possible we exploited the full set of information contained in the different CO transitions that we have observed for each source. 

\begin{table*}
{\footnotesize
\caption{CO flux measurements and luminosity ratios}
\label{tab:2}
\begin{tabular}{l c c c c c c c c c c}
\hline\hline
Source & Redshift$^{1}$ & $\Delta v^2$ & $I_{\rm CO[1-0]}$  & $I_{\rm CO[2-1]}$  & $I_{\rm CO[3-2]}$  & $I_{\rm CO[5-4]}$ & f$_{\rm 1.3mm}^4$  & R$_{\rm 21}$ & R$_{\rm 31}$ & R$_{\rm 51}$\\

 &  & km~s$^{-1}$ & Jy~km~s$^{-1}$  & Jy~km~s$^{-1}$  & Jy~km~s$^{-1}$  & Jy~km~s$^{-1}$ & mJy & & & \\
 \hline
4171    &   1.4652 & 600 &     0.178$\pm$0.039&  0.65$\pm$0.08   & 0.76$\pm$0.17 & 1.20$\pm$0.23 & 0.14$\pm$0.15 & 0.92$\pm$0.23 & 0.47$\pm$0.15 & 0.27$\pm$0.08 \\
16000   &   1.5250 & 250 &     0.240$\pm$0.027&  0.46$\pm$0.06 & 0.58$\pm$0.14 & 0.69$\pm$0.16 & 0.44$\pm$0.14 &  0.48$\pm$0.08 & 0.27$\pm$0.07 & 0.12$\pm$0.03\\
17999   &   1.4138 & 575 &      --                           &  0.58$\pm$0.06 & 0.97$\pm$0.18 & 1.28$\pm$0.17 & 0.30$\pm$0.11 & & & \\
21000   &   1.5213 & 525 &     0.161$\pm$0.026 &  0.66$\pm$0.07 & 0.83$\pm$0.16 & 1.49$\pm$0.32 & 0.84$\pm$0.36 & 1.02$\pm$0.20 & 0.57$\pm$0.14 & 0.36$\pm$0.10\\
Average$^3$ &  --  & --            &  0.193$\pm$0.018 & 0.59$\pm$0.04  & 0.73$\pm$0.09 & 1.12$\pm$0.14 & --  & 0.76$\pm$0.09 & 0.42$\pm$0.07 & 0.23$\pm$0.04\\
\hline
\end{tabular}\\
Notes:\\
1: Redshifts are taken from D10.\\
2: Velocity range for the flux measurements.\\
3: Average of the 3 BzK galaxies with all four transition measurements (BzK-4171, 16000 \& 21000). Uncertainties are standard errors of the mean. \\
4: Continuum flux density at the rest frequency of CO[5-4] (the CO fluxes listed in this table were already all corrected for the underlying continuum).\\
}
\end{table*}

 \begin{table*}
{\footnotesize
\caption{FIR SED fitting results and other physical properties}
\label{tab:seds}
\begin{tabular}{l c c c c c c c c}
\hline\hline
Source &        $<U>$ &     log L$_{\rm IR}/L_\odot$ &  log M$_{\rm dust}/M_\odot$ & f$_{\rm 1.3mm}^1$[mJy] & f$_{\rm 2.2mm}^1$[mJy] & log M$_{\rm stars}/M_\odot$ & CO FWHM[$''$]$^2$ & C$^3$\\
\hline
4171      &     16.4$\pm$4.2 &12.00$\pm$0.03 & 8.70$\pm$0.10 &         0.35$\pm$0.05 &        0.06$\pm$0.01 & 10.60 & 1.34 & 0.12 \\
16000     &      4.9$\pm$1.3 &11.85$\pm$0.03 & 9.08$\pm$0.10 &         0.52$\pm$0.08 &        0.10$\pm$0.02 & 10.63 & 1.28 & 0.07\\
17999     &     13.2$\pm$2.9 &12.06$\pm$0.03 & 8.85$\pm$0.09 &         0.48$\pm$0.06 &        0.09$\pm$0.02  & 10.59 & 0.75 & 0.11\\
21000     &     15.7$\pm$3.3 &12.32$\pm$0.03 & 9.03$\pm$0.09 &         0.66$\pm$0.19 &        0.12$\pm$0.04  & 10.89 &  0.72 & 0.18\\
\hline
\end{tabular}\\
Notes:\\
1: Fluxes predicted from the SED fitting, computed at rest frequencies of CO[5-4] and CO[3-2].\\
2: Reported from D10.\\
3. Clumpiness measured in the HST-WFC3 F160W images (Sect.~\ref{sec:clumpiness}).
}
\end{table*}

\subsection{IRAM PdBI}\label{sec:meas1}

Consistently with the CO flux values reported in D10, all CO measurements reported in this paper were extracted using circular Gaussian models\footnote{A circular Gaussian model has only one additional parameter compared to a PSF model, i.e. the size FWHM, and hence is best suited for somewhat resolved measurements with limited S/N ratios as in our case.}  for the 
sources and  fitting the visibilities in the $uv$ space using {\it uvfit} under MAPPING in GILDAS, with fixed CO sizes as given in Fig.~4-right of D10 (see also Table~\ref{tab:2} here). These CO sizes are consistent, on average, with the optical sizes measured with the Hubble Space Telescope (HST; D10). Given the lack of high spatial resolution configurations in the new observations of CO[3-2] and CO[5-4] (Table~\ref{tab:1}), we cannot improve on those size measurements, but at the same time  the measured fluxes of these high-J transitions are rather insensitive to the exact Gaussian full width at half maximum used for extraction (which is always substantially smaller than the synthesized beam).
 
To define the CO positions and velocity widths for the high-J CO observations we proceeded iteratively. Starting from the IRAM observations of CO[2-1] from D10 that  already have fairly high S/N and thus fairly small positional errors, we collapsed the data cube using velocity ranges defined from CO[2-1], to obtain single channel datasets of CO[3-2]
and CO[5-4] for each sources. We fitted the $uv$ data with an appropriate circular Gaussian model with free spatial centering, determining the position 
and the error of the CO emission that we could compare to the CO[2-1] positions with their uncertainties. In all cases the high-J CO positions were in agreement 
with the old CO[2-1] ones, except for the case of BzK-4171 in which a small offset was suggested. In that case we defined a new best CO position by combining the independent measurements in all transitions weighting with their noise, resulting in a small offset of 0.2$''$ to the North (2$\sigma$ significance) with respect to the old CO[2-1] position, comparable to its 1-$\sigma$ uncertainty (notice in Fig.2 top-left in D10 how the CO contours were slightly offset South respect to the HST imaging of the galaxy, a likely effect of noise that we are now correcting).  We then extracted CO spectra for all transitions using 25~km~s$^{-1}$ channels, at fixed spatial positions. We produced  averages of 
the CO[2-1], CO[3-2] and CO[5-4] spectra using the global S/N in each transition to determine its weight and inspected them, to verify if the velocity ranges  defined in D10 were still the most accurate definition of the spectral windows. This was the case for all sources, except again BzK-4171 where we reduced the velocity range by excluding 2 channels in the blue side and 1 in the red side, resulting in a 12\% smaller spectral range of extraction, as no sign of a signal was present at these edges when combining together the three CO transitions. By reiterating with the new channels the spatial positioning check for BzK-4171 we verified that the procedure had converged.

We caution here that while this procedure allows us to exploit all available information to provide highest fidelity measurement of CO SLEDs
within the chosen velocity range, which is determined by the CO[2-1] observations primarily and also by CO[3-2] and CO[5-4] consistently, we cannot rule out that some CO signal exists in broader components extending outside the velocity ranges where our measurements are
obtained. Such broad components are sometime seen in starburst like galaxies (Riechers et al 2011) and appear to be more prominent 
in low-J transitions, hence could alter the total CO SLEDs. Our SLEDs thus neglect such putative high velocity wings, should they be present in our targets.

The final CO spectra for all $J=2,3,5$ transitions for the 4 BzK galaxies are shown in Fig.~\ref{fig:do3}, their weighted average is in Fig.~\ref{fig:doWei}.
We also produced images of our sources in the CO[3-2] and CO[5-4] transitions (Figs.~\ref{fig:co32}~and~\ref{fig:co54}) using a natural weighting scheme.
Images and spectra were created after estimation (and subtraction) of the continuum level which is mostly relevant for CO[5-4], as described below (Section~\ref{sec:xVLA}).
All CO fluxes and their uncertainties are reported in Table~\ref{tab:2}. We also compute and report in the table CO luminosity ratios, expressed as $R_{\rm J1\ J2} = I_{\rm CO[J1\ J1-1]}/I_{\rm CO[J2\ \ J2-1]}\times (J2/J1)^2$. These are  equivalent to brightness temperature ratios 
 between transition $J1$ and $J2$  only in the limit of small Rayleygh-Jeans corrections (see Section 2.6 in Genzel et al 2010).
The final coordinates of extraction are in Table~\ref{tab:1}.  All measurements of CO[3-2] and CO[5-4] result in at least $S/N>4$, and half of them have $S/N>5$. The noise structure of the data  is not fully homogeneous though, and the CO maps  in some cases reveal non gaussian departures corresponding to negative peaks.
However, performing measurements over fixed velocity and spatial positions, largely determined from CO[2-1], can help minimize the impact of these effects given the high significance of the detection of this line in each source, i.e. we are not chasing high-$\sigma$ noise peaks in the data cubes.  We also emphasize that even non-detections or lower S/N measurements than presented here would have provided meaningful constraints on the higher excitation part of the SLED because our deep observations would comfortably detect these transitions, should they be thermally excited.

\subsection{VLA}
\label{sec:xVLA}

For the VLA data, which are the combination of datasets obtained with different correlators over the years, we did not fit the observations in $uv$ space to measure fluxes (because too complex operationally) but produced images, using a natural weighting scheme, based on the velocity information derived from IRAM, and performed flux measurements directly on the images. We could not exactly match the 25~km~s$^{-1}$
velocity channel sampling we used at IRAM (corresponding to 3.5--4~MHz at 45~GHz) due to the poorer spectral resolution of the older VLA correlator (50~MHz channels), however in all cases we could cover fully the velocity extent of the line, sometimes with up to 50~km~s$^{-1}$ wider bandwidths (which  could help alleviating problems with eventual broader CO[1-0] components, discussed in the previous section).

We then proceeded to measure CO[1-0] fluxes in the resulting images by fitting the same circular Gaussian models used for higher-J CO, but this time with 
{\it galfit} (Peng et al 2010) and using the synthesized beam as the PSF. 
Uncertainties were estimated by measuring the rms of the pixel values in the images, which is appropriate for the noise of point sources,  and scaling these values up by the ratio of the fluxes recovered for the sources 
for Gaussian  versus  point-like extractions, which was always $<15$\%. Our VLA flux measurements of CO[1-0] are reported in Table~\ref{tab:2}. 

These measurements are consistent within the errors with those reported in Aravena et al (2014), however they are not exactly the same and have somewhat 
smaller uncertainties. This is due to the fact that by using prior positional, velocity and size information from the higher-J emission we could reduce some
of the uncertainties, while also obtaining measurements that are directly comparable to those at higher-J, and also to the fact that we incorporate  the
older VLA data. The total VLA integration time on the three BzK galaxies amounts to 259~hours, after accounting for primary beam reduction (a large part of the data was obtained for the simultaneous observation of GN20 in the CO[2-1] transition; Hodge et al. 2013).

   \begin{figure}
   \centering
   \includegraphics[width=9cm]{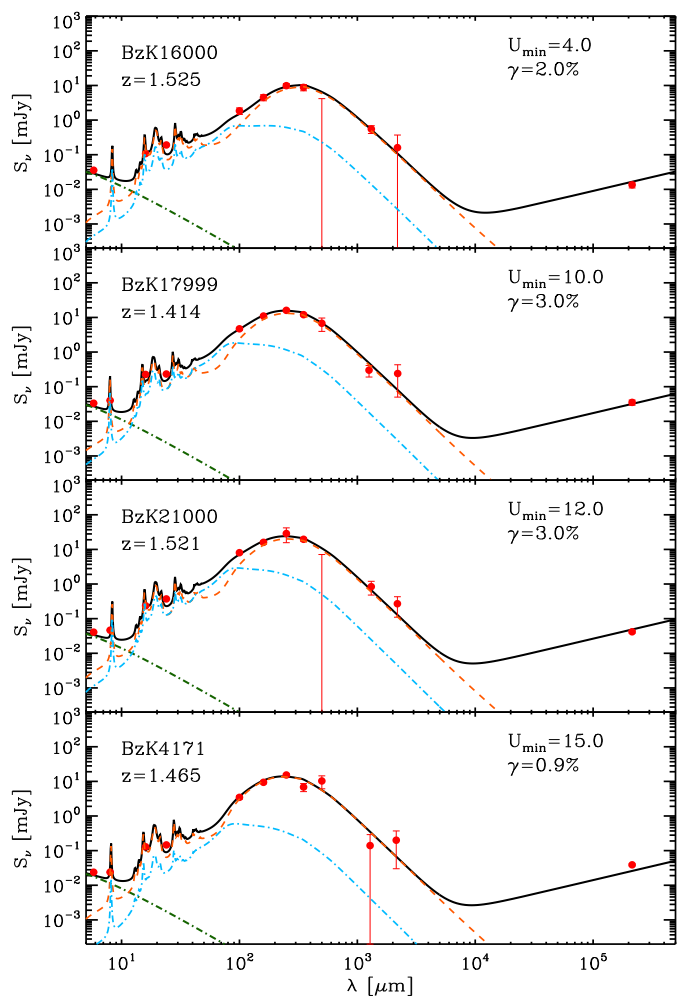}
   \caption{ IR SEDs of our sources, fitted with Draine \& Li 2007 models, extended in the radio assuming the local radio-IR correlation. Observed wavelengths are plotted. Orange dashed and cyan dot-dashed lines show separate contributions of starlight and emission from diffuse ISM component and ''PDR'' component, respectively. The stellar component that is shown with a green dot-dashed line.}
              \label{fig:seds}%
    \end{figure}
   \begin{figure*}
   \centering
   \includegraphics[width=18.5cm]{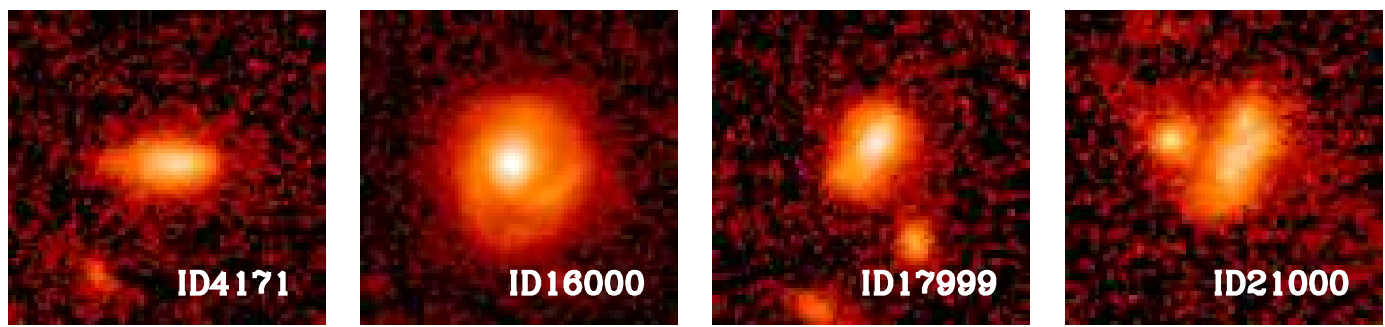}
   \caption{HST WFC3 cutouts in the F160W band from public CANDELS imaging (Koekemoer et al 2011; Grogin et al 2011). The cutouts are 5$''$ on a side 
   and are shown with logarithmic stretching to improve the visibility of details. The blob to the North-East of BzK-21000 is an unrelated object (North is up and East to the right). For comparison, Fig.~5 in D10 shows 3-color HST+ACS imaging of the galaxies.
			   }
              \label{fig:wfc3}%
    \end{figure*}

\subsection{Continuum subtraction and derivation of $<U>$}
\label{sec:U}

Given the rapidly rising dust continuum emission with frequency in the Rayleigh-Jeans tail sampled by our data, higher-J CO emission lines have increasingly 
lower equivalent width, and require a careful continuum subtraction. This is totally negligible for CO[1-0] and CO[2-1], marginally important for CO[3-2], and 
relevant basically only for CO[5-4].  We obtained a first estimate of the continuum level in the data by averaging all velocity ranges outside the detected
CO emission lines (Tab.~\ref{tab:2}; this are slightly different but consistent with what reported in Table~1 of Magdis et al 2012). Only upper limits were obtained at 2mm below the CO[3-2] transitions, while some S/N$\sim$2--4 results were found at 1.3mm. 
We then refined these low-S/N measurements by exploiting the full IR spectral energy distribution in these galaxies. We use the latest determination of the  Herschel fluxes updating measurements shown in Magdis et al (2012), based on a deblending technique that will be described in forthcoming papers (Daddi et al in preparation; Liu et al in preparation), and fit the SEDs using Draine \& Li (2007) models, following the approach in Magdis et al (2011; 2012). Results are shown in Fig.~\ref{fig:seds}. We computed the best fitting values for the 1.3mm and 2.2mm continuum flux densities and their uncertainties from the fit (Table~\ref{tab:seds}). Comparing  to the measurements  we found general consistency within the errors. We combined both (fit and measurement) information at 1.3mm weighting using their errors, to define the best continuum value for subtraction from the CO[5-4] flux. The associated uncertainties are much smaller than those we obtained from the direct estimate of the continuum to either side of the CO[5-4] line in the IRAM data alone. At 2.2mm we subtracted the flux predicted by the fit as the only available continuum estimate. This correction is small ($\simlt5$\%) in all cases. Even if the formal measurements at 2.2mm tend to be in excess of the best fit SED (Fig.~\ref{fig:seds}), we verified  that the SED fits are consistent within the errors also when stacking the four sources. 

These SED fits also allow us to obtain updated estimates of the bolometric L$_{\rm IR}$, $M_{\rm dust}$  and thus mean radiation intensity $<U> = 1/P_0\times L_{\rm IR}/M_{\rm dust}$ for our targets ($P_0=125 L_\odot/M_\odot$, so that $<U>$ is dimensionless; Draine \& Li 2007). These quantities are 
reported in Table~\ref{tab:seds} and will be used in the remainder of the paper.  They agree within the uncertainties with values published in Magdis et al 2012
(their Table~2). We recall that $<U>\propto T_{\rm dust}^{4+\beta}$ and, consistently with the findings in Magdis et al (2012), these $<U>$ values correspond to dust temperatures in the range of 30--35~K for modified Black-Body fits with $\beta\sim1.5$. The typical uncertainties on $<U>$, given the quality of our multi-wavelength dataset, are of order of 0.1~dex (20--30\%). We also show in Fig.~\ref{fig:wfc3} the HST Wide Field Camera~3 (WFC3) images of our targets. 
{ The F160W imaging, together with the UV rest-frame imaging (see Fig.~5 in D10) reveal morphologies which are reasonably regular especially in the rest frame optical but characterised by massive clumps, fairly consistent with the typical clumpy disk galaxies at high redshifts (Genzel et al 2006; Bournaud et al 2008; Forster-Schreiber et al 2009).}
   
  \begin{table}
{\footnotesize
\caption{Average CO SLED of the (U)LIRGs from P12 used in this work}
\label{tab:ulirgs}
\begin{tabular}{l c c c}
\hline\hline
 J       & $I_{\rm CO}^1$      &      $\Delta I_{\rm CO}^2$    & N$_{\rm bands}^3$ \\
 \hline
              1     &    1.00     &    0.00       &      14  \\
              2      &   3.59     &    0.22         &     13 \\ 
              3       &  7.11      &   0.53          &    14\\
              4      &  12.4      &   1.40         &      6\\
              5      &  12.2      &   1.90         &      0\\
              6      &  10.6      &   2.40          &    14\\
\hline
\end{tabular}\\
Notes:\\
1: Fluxes of individual galaxies have been normalised to CO[1-0].\\
2: The error is the r.m.s. dispersion of individual measurements divided by square root of number of galaxies in the sample.\\
3: Total number of measurements available for the transition from P12. Lacking a direct measurement, we have used 
an estimate based on LVG modelling of the other available transitions.\\
}
\end{table}

\subsection{Comparison samples of local spirals and (U)LIRGs}
\label{local}

We use a representative sample of (U)LIRGs taken from 
from P12, for which the contribution from star formation is $L_{\rm IR}>2\times10^{11}L_\odot$. 
{ These authors do not observe CO[5-4] directly but their sample includes sources
with a measurement of CO[6-5], sometimes including CO[4-3], and virtually always  CO[3-2], CO[2-1] and CO[1-0], see Table~\ref{tab:ulirgs}. For these  galaxies we used
single component LVG models
(see Section~\ref{lvg} for more details on the models)  to interpolate between CO[6-5] and lower-J transitions, excluding however CO[1-0] and in some cases CO[2-1], depending on the fit (consistent results were found with two component LVG fitting when including the lowest-J transitions). We visually inspected all fits and retained 14 galaxies 
where the LVG fitting 
solutions  appear to un-ambiguously return estimates for the CO[5-4] fluxes.} We construct an average CO SLED of local (U)LIRGs by averaging all available measurements of the 14 galaxies\footnote{
IRAS00057+4021,              IRAS02483+4302,                                                                               IRAS04232+1436,             IRAS05083+7936-VIIZw031,   
IRAS05189-2524,                                                                               
IRAS08572+3915,                                                                               
IRAS09320+6134-UGC5101,     
IRAS10565+2448,             
IRAS12540+5708-Mrk231,                                                                        
IRAS13428+5608-Mrk273,                                                                        
IRAS15327+2340-Arp220,                                                                        
IRAS16504+0228-NGC6240,                                                                       IRAS17208-0014, and                                                                           
IRAS23365+3604.   
}  from P12 and using the modelled value for lacking measurements. The average (U)LIRGs CO SLED derived in this way is reported in Table~\ref{tab:ulirgs}.

For spiral galaxies we use archival Herschel observations of the CO[5-4] emission in five Kingfish galaxies\footnote{
NGC3351,
NGC3521,
NGC4736,
NGC6946 and 
NGC7331.
}
 using the SPIRE FTS spectrometer. { Observations were performed in mapping mode and we used data from the central bolometer (36$''$), covering the nucleus and inner disk of the galaxies.} Details of the measurements will be presented
in Liu et al (2015, in preparation). We also use measurements from HERACLES (Leroy et al 2009) of CO[2-1]. We determine the aperture correction of the Herschel  CO[5-4] measurements using the IRAS imaging. In addition, we use data for the local spiral IC342 from Rigopoulou et al. (2013) and the MW.

\section{Results}\label{sleds}

\subsection{CO SLEDs of normal $z=1.5$ galaxies}

\begin{figure}
   \centering
   \includegraphics[width=9cm]{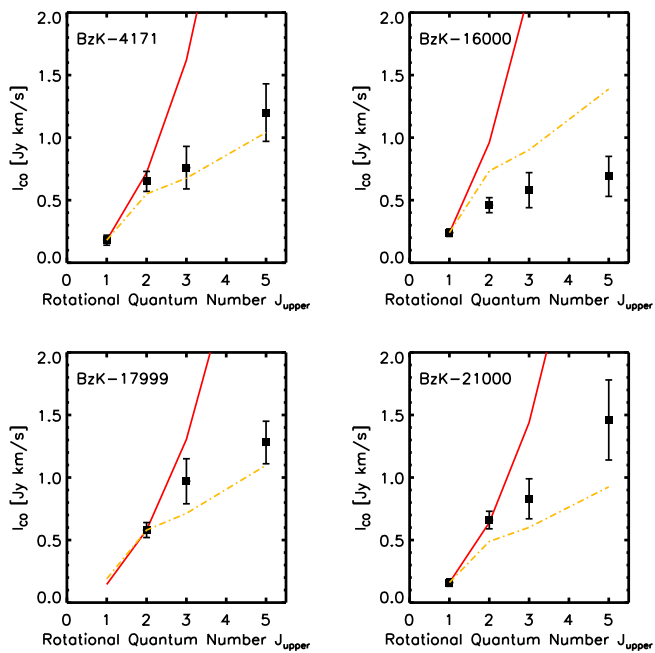}
   \caption{CO Spectral Line Energy Distributions (SLEDs) for the 4 BzK galaxies in our sample, including IRAM PdBI data over CO[2-1], CO[3-2] and CO[5-4], and VLA measurements of CO[1-0] for 3 galaxies, based on Aravena et al. (2014). The dot-dashed line show the average SED for comparison, emphasizing excitation variations within our sample. The solid line corresponds to a constant brightness temperature.
			   }
              \label{fig:h4}%
    \end{figure}

The multi-transition CO SLEDs of our 4 targets are shown in Fig.~\ref{fig:h4}, including CO[1-0] (except for BzK-17999), CO[2-1], CO[3-2] and CO[5-4]. 
We have also averaged the fluxes for the four CO transition for the three BzK galaxies with full SLED observations. Results are shown in Fig.~\ref{fig:hmed} and the fluxes are in Table~\ref{tab:2}.
Comparing to the trend expected for thermally excited CO emission (red curves, normalized to the lowest-J transition available, with fluxes scaling $\propto J^2$), we see that generally CO[2-1] 
is only weakly sub-thermally excited. For two galaxies (BzK-21000 and BzK-4171) the CO[2-1] to CO[1-0] brightness temperature ratio $R21$ (Table~\ref{tab:2}) is consistent with unity, while BzK-16000 is already significantly sub-thermally excited at CO[2-1]. As a result, the average $R21$ for our sample
is $0.76\pm0.09$, moderately sub-thermally excited. We note that this is formally consistent with the value of 0.86 that we derived in Dannerbauer et al. 2009 
and used to correct CO[2-1] fluxes to CO[1-0] in D10 and subsequent papers from our group (Daddi et al 2010b; Magdis et al 2011; 2012; Sargent et al 2014; etc).

Conversely, CO[3-2] and CO[5-4] are both significantly sub-thermally excited in all sources. For the $J=3$ transition we measure average R$_{31}=0.42\pm0.07$, 
with ranges from 0.27 to 0.57 for individual sources. It is clear that estimating the CO[1-0] flux and thus total CO luminosities of $z=1.5$ galaxies from CO[3-2]
is  less accurate than using CO[2-1]. However we find that the typical assumption  of R$_{31}\sim0.5$ which is adopted in the literature (Tacconi et al 2010; 2013; Genzel et al 2010; etc) is consistent with our data on average. Globally, the 4 sources behave quite similarly to BzK-21000 which was first discussed in Dannerbauer et al. (2009; see also Aravena et al 2010; 2014), and in particular their SLEDS from CO[2-1] to CO[3-2]  closely resemble  the MW 
(Fixsen et al 1999) and presumably the same holds at CO[1-0] although the MW Fixen data has a very large error there (see Fig.~\ref{fig:h4}).

However, it is  clear that when going to the CO[5-4] transition the average CO SLED deviates strongly from that of the MW and other local
spirals: the CO[5-4] emission of the 
BzK galaxies is about 4 times higher on average than what would be expected normalizing the MW SLED to the $J\leq3$ transitions. We see CO[5-4] flux
in excess of MW expectations in all cases, although for BzK-16000 the excess is the lowest, only of a factor of two. The average R$_{51}$ is $0.23\pm0.04$, with
ranges for individual sources going from 0.12 to 0.36, or about a factor of 3. This suggests that extrapolating CO[1-0] fluxes from CO[5-4] for this population  would likely lead to uncertainties of the order of factors of two. 

Globally, the BzK galaxies have CO SLEDs that are less excited than those of local  (U)LIRGs. P12 (U)LIRGs have on average CO[3-2]/CO[2-1]  ratio  $>0.8$ for their sample, compared to 0.52 in our case. Similarly,  we find that starbursting local (U)LIRGs have higher CO[5-4]/CO[2-1] flux ratios by a factor of 1.8.

Comparing to SMGs, for which Ivison et al. 2011 report R$_{31}=0.55\pm0.05$ and Bothwell et al 2013  R$_{31}=0.52\pm0.09$, we see that the normal MS galaxies at $z=1.5$ appear to have somewhat less
excited CO SLEDs (R31$=0.42\pm0.07$). Similarly, Bothwell et al 2013 reports R$_{51}=0.32\pm0.05$ which is also higher than our average value
(0.23$\pm$0.04), while the Spilker et al 2014 spectrum of gravitationally lensed sources discovered with the South Pole Telescope (SPT) has even higher ratios,
implying that the CO[5-4]/CO[1-0] flux ratios for SMGs are higher by 40\% or more on average.  This is confirmed by the comparison of the global SED in Fig.~\ref{fig:hmed}, which shows a higher excitation for the SMGs at $J\geq3$. We note  from Fig.~\ref{fig:hmed} that, 
in comparison, the CO SLED of SMGs is intermediate in excitation strength between that of local (U)LIRGs and $z=1.5$ BzKs. This  supports the notion  that SMGs are not purely SBs but probably a fair mix of MS and SB populations (see e.g. Rodighiero et al 2011), although the higher excitation of the SPT sample might also be connected to their  higher average redshift (see next sections).
There are individual SMGs with CO excitation  comparable to that of the BzK galaxies (e.g., Riechers et al 2011).

\subsection{Comparison to theoretical predictions of CO SLEDs}

\begin{figure}
   \centering
   \includegraphics[width=9cm]{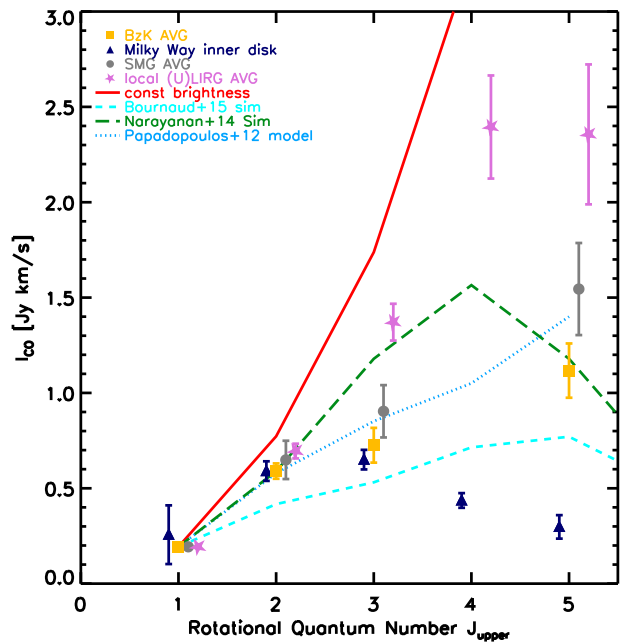}
   \caption{Comparison of the average SLED of BzK galaxies  to the MW SLED (Fixsen et al. 1999), the average of SMGs (Bothwell
   et al 2013; the Spilker et al 2014 SLED is similar but higher at $J>5$) and the average (U)LIRGs SLED derived in this paper using measurements from P12. All SLEDs are normalized to the CO[1-0] of the average BzK galaxy SLED, except the MW which is normalised using CO[2-1].
   The   results from the toy model of P12 (see text) and the numerical simulations of B15 and Narayanan \& Krumholz (2014) are also shown.
			   }
              \label{fig:hmed}%
    \end{figure}

In Fig.~\ref{fig:hmed} we compare the average SLED of the BzK galaxies with several predictions for high redshift disks from theoretical modeling, which are all in fair agreement 
with the data, at least in a qualitative sense. These models correctly predict an enhancement over the MW CO SLED although complete agreement with the average BzK data is not achieved by any model at this stage. 

P12 (blue dotted line) provide predictions for high-z BzK galaxies analogous to our sources based on simple reasoning and assuming a scaling of the CO excitation depending on the SFE or equivalently the fraction of dense gas (assuming a universal SFE per dense molecular gas mass). In their interpretation, the high-J CO
emission is heated by turbulence and/or cosmic rays in the dense gas (e.g., Papadopoulos et al 2014).
Their
expected SLEDs is not far from the observed one, but rises more rapidly than the data at the highest transitions. It seems plausible that full agreement with the 
observations could be obtained by tuning some of the free parameters for which they could only provide guesses. 

  \begin{figure*}
  \centering
   \includegraphics[width=18.5cm]{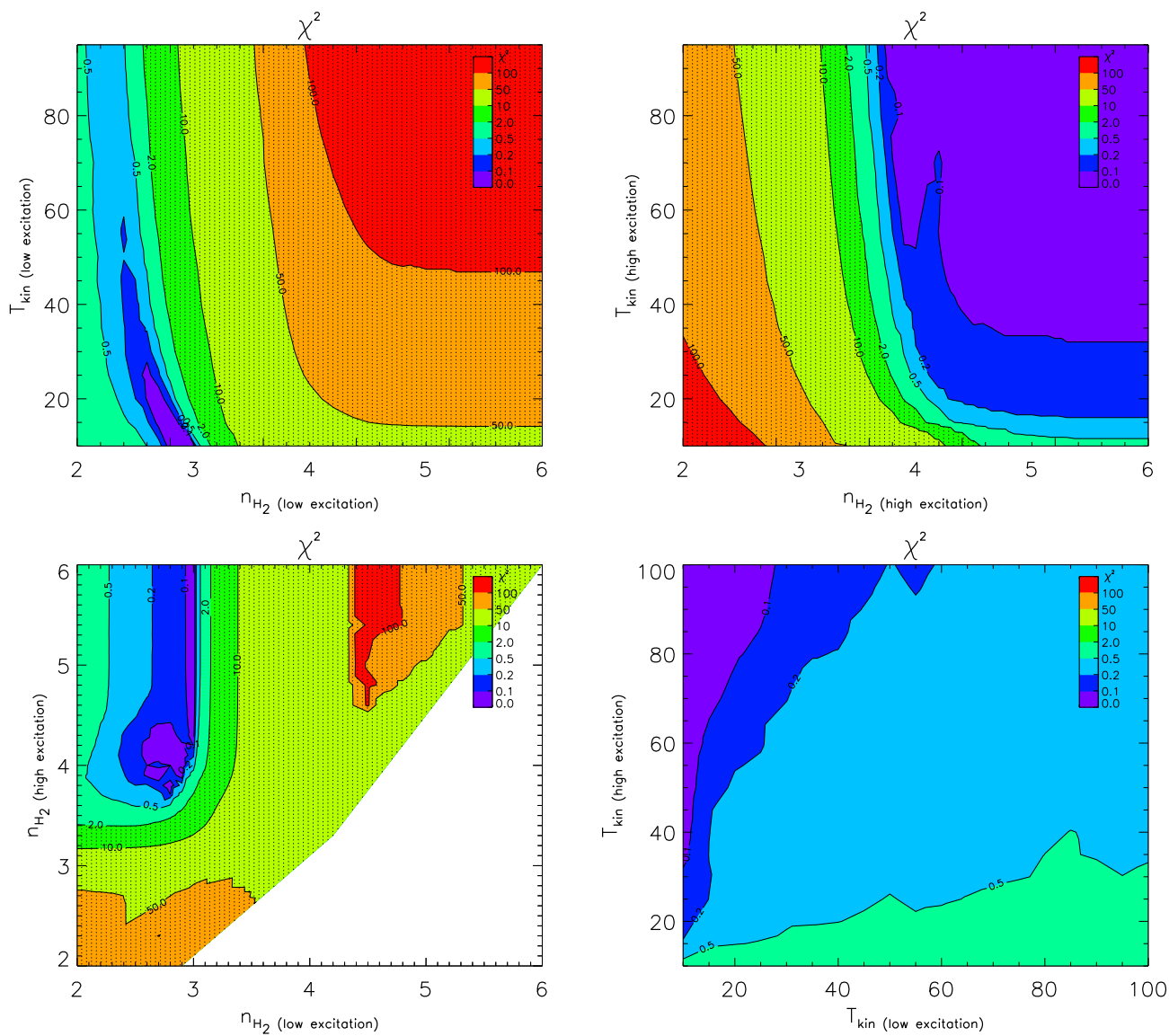}
   \caption{Total $\chi^2$ maps versus temperature and density for the fit of the BzK average CO SLED with two components (a low excitation one and a high excitation one, the latter defined to peak at higher J).  The   color scale reflects the $\chi^2$ level: blue/light green regions correspond to acceptable fits. For a choice of variables in each panel the minimum $\chi^2$ is computed by marginalising over the other two free parameters.
			   }
              \label{fig:chimap}%
   \end{figure*}

   \begin{figure*}
   \centering
   \includegraphics[width=9cm]{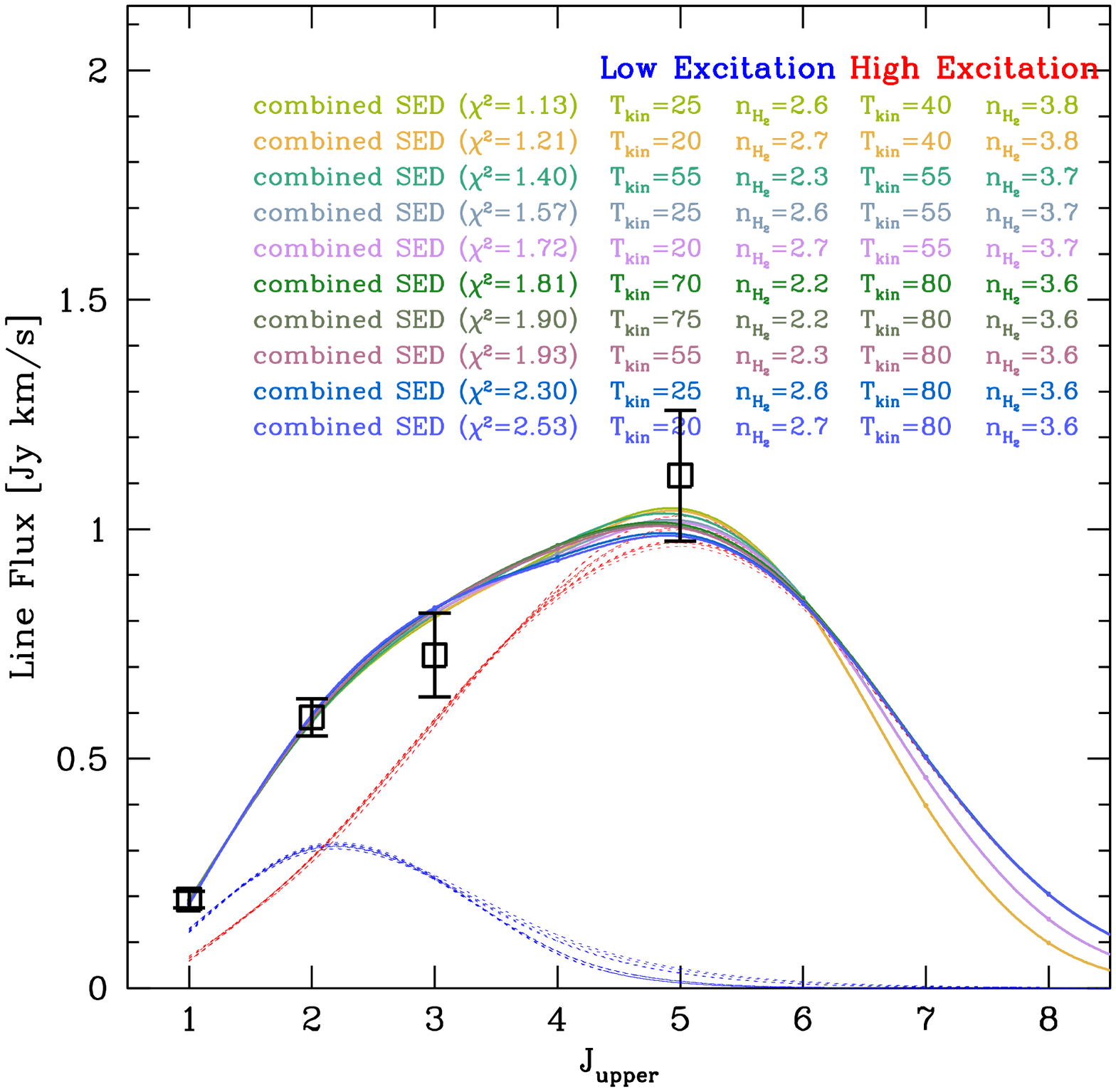}
   \includegraphics[width=9cm]{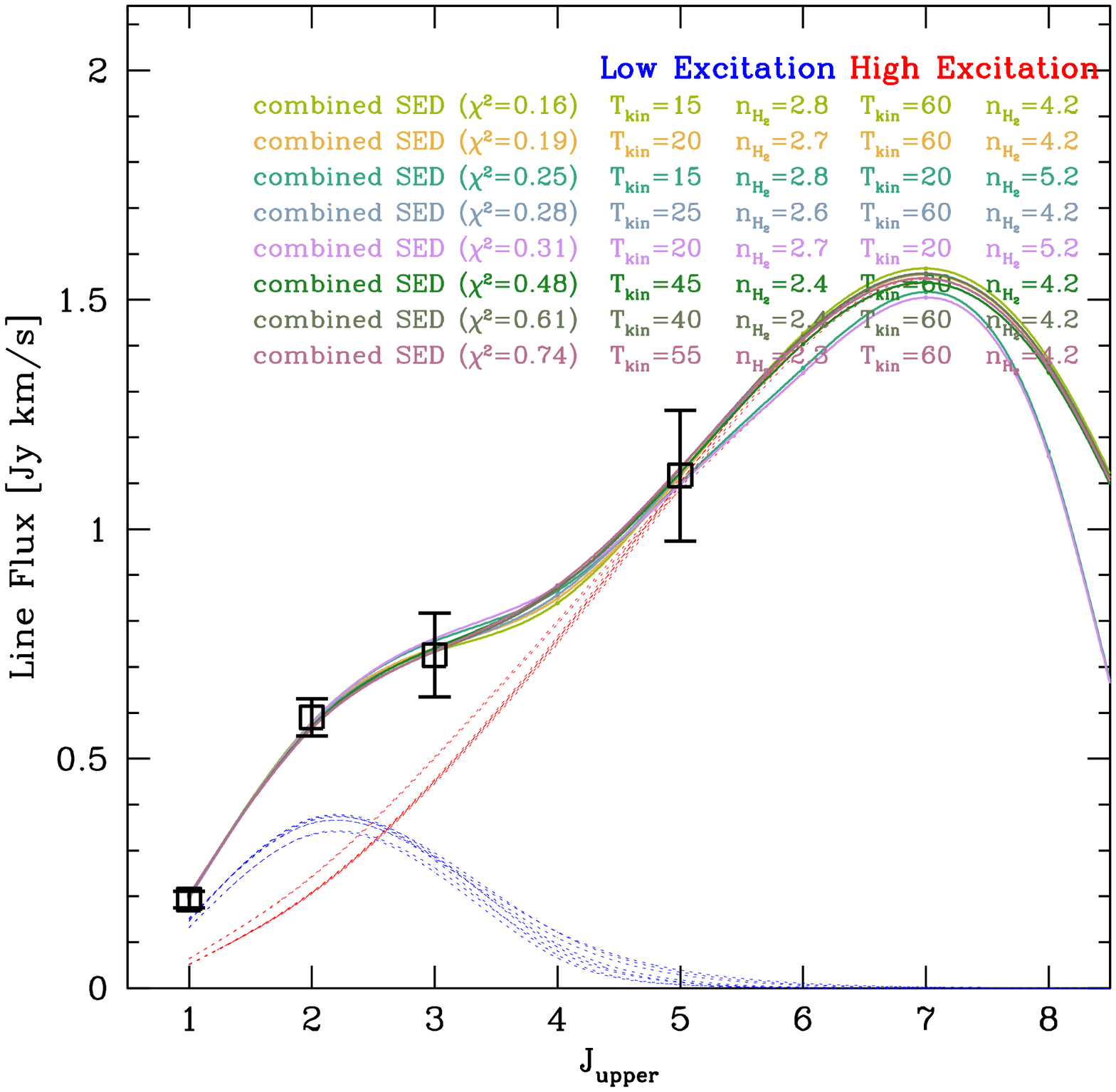}
   \caption{Examples of LVG fit to the average CO SLEDs of BzK galaxies. These particular solutions (among the large suite of acceptable ones) were chosen to have a cold component peaking at $J\sim2$ and a warmer/denser component peaking at $J\sim5$ (left panel) and at $J\sim7$ (right panel). The captions show details about the physical parameters of each solution. Solutions peaking at higher $J$ obviously exist and can fit the data also well. 
			   }
              \label{fig:lvg}%
   \end{figure*}

More recently, Narayanan \& Krumholz (2014) have developed a general model to predict CO observations over different transitions by combining numerical simulations and radiative transfer calculations. In their case  the overall excitation is 
regulated by the star formation surface density ($\Sigma_{\rm SFR}$; { a similar behaviour is predicted by Popping et al 2014}). Based on the typical $\Sigma_{\rm SFR}\sim1$~M$_\odot$~yr$^{-1}$~kpc$^{-2}$ for BzKs and normal $z=1$--2 disk galaxies (Daddi et al 2010b; Genzel et al 2010), we can use their predictions to compare to BzK galaxies (green line in Fig.~\ref{fig:hmed}). The Narayanan \& Krumholz (2014) model  correctly predicts enhanced CO[5-4] emission compared to the MW (Fig.~\ref{fig:hmed}), although with 
an overall shape that is somewhat different than that of the BzKs, rising steeply towards CO[4-3] and resulting in too large CO[3-2]/CO[2-1] ratios.
It appears that the CO SLED of BzK galaxies has an unexpected shape, fairly flat and MW-like up to CO[3-2] 
and rising quite rapidly afterwards, giving peculiarly high CO[5-4] (or CO[4-3]) to CO[3-2] ratios (the agreement with a MW shape up to $J=3$ is correctly anticipated by P12). 

Finally, we compare to the predictions of CO SLEDs for high-$z$ disk galaxies by B15, based on  very high resolution AMR numerical
simulations. Also in this case the models predict enhanced CO[5-4] emission compared to the MW, although the  discrepancy here is that the expected SLED is less excited than observed, with expected R$_{51}\sim0.14$. In the B15 simulations  the high-J excitation component of the
CO SLED is attributed largely to the giant clumps while the rest of the diffuse SF in the galaxy contribute a low-J excitation component. In this respect, the lower than observed excitation of CO can be explained by an underestimate of the gas fraction compared to the real galaxies, because a lower gas fraction will result in fewer/less massive clumps in the galaxies (see Salmi et al 2012; Bournaud et al 2012; Bournaud et al 2011). The low gas fraction is due to the fact that the
simulations start with a (reasonable) gas fraction of $\sim50$\% (albeit this is for total gas, molecular gas fraction could even be lower), but given  the lack of gas accretion in the simulations during the lifetime of the galaxy the gas fraction decreases and is less optimal. Another 
explanation for the lower R$_{51}$ in the models might be resolution effects biasing low the computation of CO[5-4] emission (B15). 
In any case, B15  successfully predict a difference in the CO SLED between local (U)LIRGs and high-$z$ disks, in qualitative agreement with observations. 

{ Overall, the B15 model provides too little flux in all lines above CO[1-0] (or, equivalently, too strong CO[1-0] emission), while the Narayanan \& Krumholz model predicts a SLED shape different from the data. The P12 model provides too much flux at $J\geq3$, but performs somewhat better that the other models.  When looking at individual sources rather than the average SLED, we can see that the BzK galaxy with highest CO excitation (BzK-21000) is  consistent with the expectations of P12. The high-J transitions of the galaxy with the lowest excitation, BzK~16000, are even less excited than the significantly sub-thermal excitation  predicted by the simulations of B15.}

\section{LVG modeling}\label{lvg}

Large Velocity Gradient (LVG) modeling is a classical tool for studying CO SLEDs and gaining insights into the physical properties of the gas,  including their kinetic temperature and density (Tkin; n; Goldreich \& Kwan 1974; Scoville \& Solomon 1974). We follow the approach described in Weiss et al (2007), Dannerbauer et al (2009) and Carilli et al (2010), using the collision rates from Flower (2001) with an ortho-to-para ratio of 3 and a CO abundance per velocity gradient of [CO]/$\Delta v = 1\times10^{-5}$~pc~(km s$^{-1}$)$^{-1}$. We compute model grids using an appropriate value of the CMB temperature for $z=1.5$ and varying the density  over $10^2$--$10^6$~cm$^{-3}$ and Tkin over 5 to 100K.

Single component models are inadequate for fitting the whole SLED of the BzK galaxies, as it is obvious  from the broad shape with a fairly flat flux level from CO[2-1] to CO[5-4]. This is true for the composite SLED and also for individual sources, except for BzK-17999 which can be formally fit by a single component but, probably, just due to the lack of CO[1-0] constraints.
On the other hand, we find that two component models are highly degenerate and a large variety of acceptable solutions 
can be found. This is  at least in part due to the intrinsic degeneracies of this kind of modeling, where increasing density or temperature affects the SLEDs in the same way at least to first order, but also due to the fact that measured fluxes continue to rise all the way to CO[5-4] and that we have no evidence of where beyond this transition the SLED begins to turn down.  In Fig.~\ref{fig:chimap} we show a map of $\chi^2$ values as a function of the density and temperature, separately for a low-excitation and a high-excitation two-component fit. It appears that a MW-like low-excitation component with log~$n/{\rm cm}^{-3}\sim$2--3 and moderate temperature Tkin$\sim20$--30K, which is comparable to the dust temperature in these galaxies (Magdis et al 2012; Sect~\ref{sec:U}), together with a denser and/or warmer component can fit the data well. Other/different viable solutions exist as well obviously, but 
good $\chi^2$ regions appear to constrain more stringently the density of both components, with low-excitation one preferring log~$n\simlt3$ and the
high excitation one having log~$n\simgt3.5$, at least when limiting to solutions with Tkin$<100$K as we have assumed. Allowing for temperatures substantially higher than 100~K in the high excitation component would reduce the upper limit on its density, as can be guessed
from the trends in Fig.~\ref{fig:chimap} (top-right).

{ Overall these results are consistent with the simulations of B15 which suggest that the high excitation component (mainly related to the giant clumps)
is denser but also warmer than the diffuse gas, and its particularly high temperature (100~K for $n=3300$~cm$^{-3}$ in a CO[5-4] luminosity weighted average of the molecular gas in B15 -- an estimate which might  still retain some {\em contamination} from diffuse gas) is key in driving its strongly peaked shape towards CO[5-4]. Their predicted values are reasonably consistent within the range of acceptable values from our modeling, although our constraints on Tkin are not very stringent and the B15 predicted density is on the low-side of the acceptable values for the high excitation component. }

In Fig.\ref{fig:lvg} we show particular LVG solutions fitting the data with reasonable $\chi^2$ values, chosen to display a peak of the low-excitation component at 
$J\sim2$ and at $J\sim5$ (left-panel) or $J\sim7$ (right panel) for the high excitation component.
Representative density and kinetic temperature values for the adopted solutions are shown in the caption of the figure. 
This kind of behavior, as in the left panel of Fig.\ref{fig:lvg}, with an overall peak of the CO SLED not far from CO[5-4] might be expected both based on the simulations of B15 and Narayanan \& Krumholz (2014). The calculations of P12 would suggest a peak at even higher-$J$, 
{ similar to the warmest among local LIRGs (e.g., Lu et al 2014)}.

LVG modelling can be used to estimate the molecular gas mass of the CO emitting galaxies, basically exploiting the constraints on the gas densities. However
this requires critical assumptions on key physical parameters which enter the calculation of the CO SLEDs at various luminosities for a given gas mass, namely: 
the disk scale height (or turbulence velocity of the medium) and the CO abundance per velocity gradient. Hence final estimates are highly uncertain, as  direct
constraints on this parameters are not very accurate. However we verified that using plausible estimates for these parameters we can match the gas masses estimates of order of 5--8$\times10^{10}M_\odot$ typically inferred for these galaxies using dynamical methods (D10), or converting from dust mass estimates (Magdis et al 2011; 2012), or  assuming a local metallicity scaling for the $\alpha_{\rm CO}$ conversion factor (Sargent et al 2014). We also find that typical LVG solutions suggest that a non negligible fraction of the total gas mass is in the more excited component, reaching in some case half of the total gas. This is consistent with B15 predictions. 

   \begin{figure*}
   \centering
   \includegraphics[width=9cm,angle=-90]{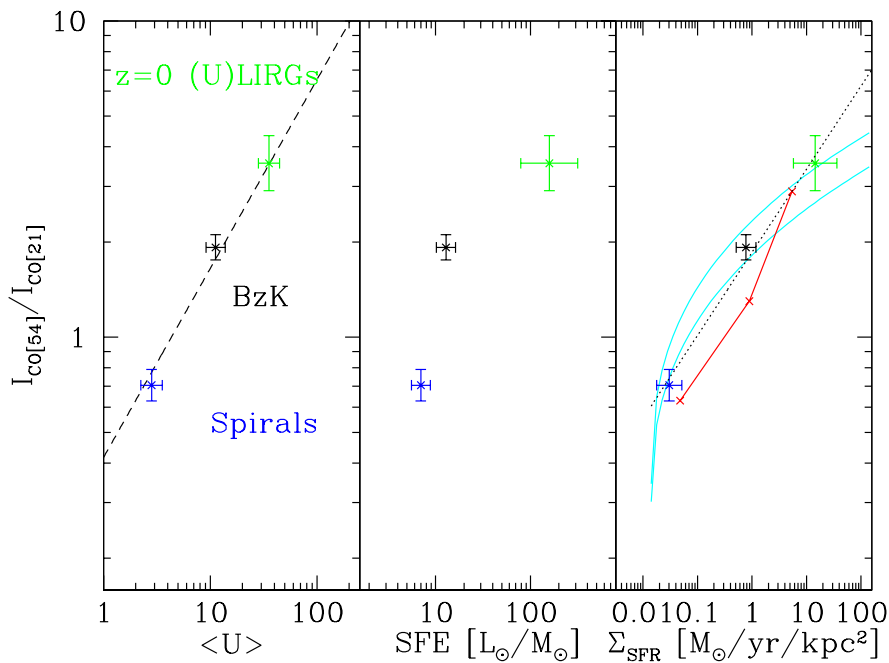}
   \caption{Comparison of the average  CO[5-4]/CO[2-1] ratio to the average radiation field intensity ($<U>$, left panel), the star formation efficiency (SFE, center panel) and the star formation surface density ($\Sigma_{\rm SFR}$; right panel), for local spirals (blue), local (U)LIRGs (green) and $z=1.5$ normal MS galaxies (black). The best fitting trend (dashed line) in the left panel is given in Eq.~\ref{eq:U}, while the
   relation in the right panel (dotted line) is in Eq.~\ref{eq:s_sfr}. The cyan trends (right panel) shows the predictions of Narayanan \& Krumholz (2014) based on their Eq.~19 and Tables~2 and 3, while in red we show the B15 simulation results.
			   }
              \label{fig:SFE}%
   \end{figure*}

%

\section{Interpretation and discussion}\label{disc}

\subsection{What is the main driver of CO excitation ?}

We have shown that CO[5-4] emission inside normal MS galaxies at $z=1.5$ is substantially excited, a result markedly different from what is seen in  local spirals 
and the MW. There are a number of reasons why this could be expected, as anticipated in the introduction to this paper, namely: the rising dust temperature in MS galaxies as reflected by the $<U>$ parameter (e.g., Magdis et al 2012), the higher fraction of dense gas and SFE (e.g., Daddi et al 2010b), the harder ionizing radiation field (Steidel et al 2014), the likely higher average gas density. Similarly, models appear to suggest different alternatives as the driver of higher excitation 
gas in high-z BzK galaxies: the SFE (P12), the $\Sigma_{\rm SFR}$ (Narayanan \& Krumholz 2014) or the warm/dense gas in the clumps
(B15). 
It is therefore relevant to try to gain insights on the different possibilities using our observations. We have collected measurements of the CO[5-4]/CO[2-1] flux 
ratios of various relevant galaxy populations from the literature and from publicly available observations. 
As anticipated in Section~\ref{sleds}, for local starbursts we find an average flux ratio of  CO[5-4]/CO[2-1] of 3.5 with an error on the average of 0.09~dex inferred from the dispersion of the measurements of the (U)LIRGs, while for local spirals we find an average ratio of 0.7 with an error of 0.05~dex. 

In Fig.~\ref{fig:SFE} we compare the average CO[5-4]/CO[2-1] of these three key populations to their average $<U>$, SFE and $\Sigma_{\rm SFR}$. 
The $<U>$ values for the four BzK galaxies were updated in this work and are shown in Table~\ref{tab:seds} with their error bars. For the spiral galaxies
we fitted their IR SEDs using Draine and Li (2007) models to the photometry provided by the Kingfish project (Dale et al 2012), 
in analogy to the procedures used for
the BzK galaxies (see Magdis et al 2011; 2012). We use here and in the following only those objects for which $<U>$ could be measured,  limiting the sample to 6 objects in total. 
For the (U)LIRGs we use the average $<U>$ value of 35 from Magdis et al (2012) and daCunha et al (2010) as only for 3 galaxies individual measurements 
are available ($<U>=      32.5$,~24.6~and~52.8 for IRAS10565, IRAS17208 and      
IRAS12112, returning an average very close to the adopted value also for this smaller subsample). {For estimating typical values for the SFE we adopt values reported in Sargent et al. (2014). The adopted values, reported in the figure, depend on our current understanding of the determination of total gas masses. This is based on CO measurements  with metallicity dependent $\alpha_{\rm CO}$ for local spirals and dynamical derived  $\alpha_{\rm CO}$ for local ULIRGS, while for BzK galaxies they are derived by concordant estimates based on CO luminosities  and dust mass measurements. Values are constrained to within typical ranges less than a factor of 2, but we adopt a larger error for ULIRGs to account for substantial spread in the population. 
The $z=1.5$ BzK galaxies have only $\simlt2$ times higher SFE than local spirals, consistent with the small tilt of the KS relation for MS galaxies, while (U)LIRGs have about 10 times the SFE of BzK galaxies.} We note that the average $L_{\rm IR}$ in the (U)LIRG sample used here is consistent with the one in Sargent et al (2014).  For $\Sigma_{\rm SFR}$ we use sizes measured by Leroy et al (2008) for spirals, 
for (U)LIRGs we use measurements from Kennicutt (1998), Scoville  (1994), and Downes \& Solomon (1998),
and for BzK galaxies we use the optical sizes (D10; Table~\ref{tab:seds}).

  \begin{figure}
   \centering
   \includegraphics[width=9cm]{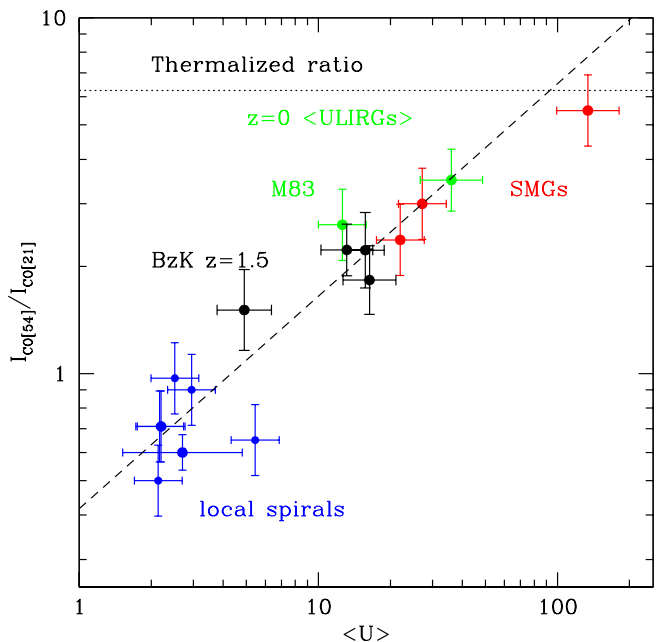}
   \caption{The ratio of CO[5-4] to CO[2-1] flux is shown versus the average radiation intensity $<U>$ for the 4 BzK galaxies in this study. They are compared to 
   the MW and a sample of local spirals from Kingfish (Dale et al 2012) observed with Herschel FTS (Liu et al. 2015, in preparation), to the average of (U)LIRGs from 
   P12 and the M83 starburst (Wu et al 2014), and to starburst-like SMGs. The horizontal line shows the limiting CO[5-4] to CO[2-1] flux ratio expected for the case of constant brightness temperature. The dashed line shows the best fitting trend to the data (Eq.\ref{eq:U}).			   }
              \label{fig:U}%
    \end{figure}

It is quite instructive to compare the three panels of Fig.~\ref{fig:SFE}. Good correlations of the CO[5-4]/CO[2-1]  ratio are found for the $<U>$ parameter (left panel) and the $\Sigma_{\rm SFR}$.
However, local spirals and $z=1.5$ MS galaxies have comparable SFEs (center panel), so that it is difficult to explain the stronger CO[5-4] emission in BzK galaxies 
only based on their higher SFE, e.g. also compared to the location of (U)LIRGs. The correlation between CO excitation and $\Sigma_{\rm SFR}$ agrees with predictions of the Narayanan \& Krumholz (2014) model fairly well, and also displays a similar trend to B15 predictions. 

\subsection{A sub-linear correlation of CO excitation with $<U>$}

We explore further the correlation between CO[5-4]/CO[2-1] ratio  and $<U>$ in Fig.~\ref{fig:U}, where we now show individual measurements for spiral galaxies
and BzK-galaxies. We also add here a sample of SMGs for which $<U>$ has been measured by Magdis et al (2012) and that are classified as starburst. (Recall that general SMG samples are instead likely mixed ensembles of MS and off-MS SB galaxies, e.g. Rodighiero et al 2011.) These include GN20 where CO measurements are from Carilli et al (2010), SMM-J2135 (cosmic-eye) from Danielson et al (2011) and HSLW-01 where we take CO measurements from Scott et al (2011).  The latter two are lensed sources, and we used de-lensed luminosities for this work.
We  also add  the low-luminosity starburst M83 taking measurements from Wu et al (2014). We still use an average $<U>$ value for the (U)LIRGs because of the lack of enough individual measurements. The data is best fit by the relation:

   \begin{figure}
   \centering
   \includegraphics[width=9cm]{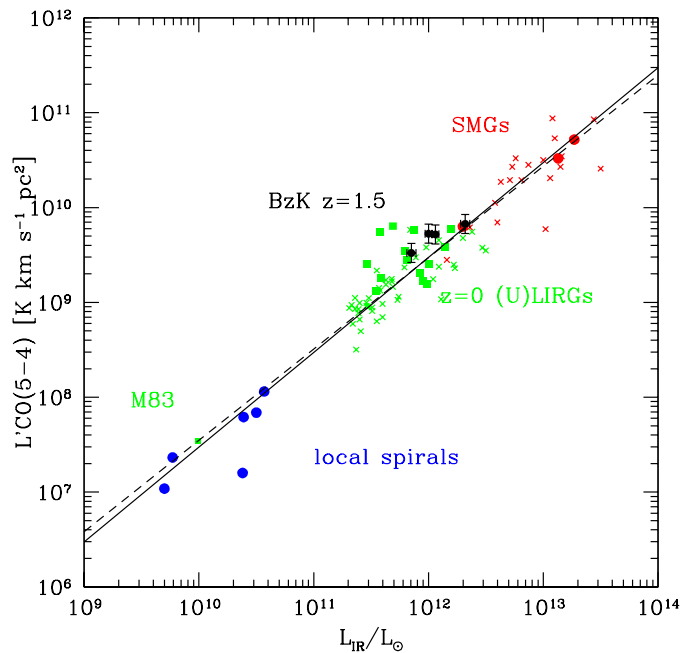}
   \caption{Correlation of the CO[5-4] line luminosity with the bolometric IR luminosity ($L_{\rm IR}$; 8--1000$\mu$m) of our 4 BzK galaxies at $z=1.5$.
   Including the sample of galaxies from Fig.~\ref{fig:SFE} (solid symbols) and additional SMGs and (U)LIRGs (crosses; see text)  we observe a virtually linear correlation 
   between the luminosities (solid line, while the dashed line is the non-linear fit), spanning about 4 orders of magnitude along both axis. The average $L'_{\rm CO[5-4]}/L_{\rm IR}$ ratio of the BzK galaxies
   is consistent with the average of that of local (U)LIRGs or high-$z$ starbursting SMGs.
			   }
              \label{fig:lir54}%
   \end{figure}

\beq
{\rm log}\ I_{\rm CO[5-4]}/I_{\rm CO[2-1]} = 0.60\times {\rm log}<U> - 0.38
\label{eq:U}
\eeq

\noindent
with a formal uncertainty in the slope of 0.07 and a dispersion in the residual of only 0.11~dex (a 30\% accuracy in predicting the CO[5-4]/CO[2-1] ratio based on $<U>$). The relation is thus substantially sub-linear. We estimate that for the CO[5-4]/CO[1-0] ratio the correlation with $<U>$  would steepen 
to a slope of $\sim0.7$ (because of the systematically rising R$_{\rm 21}$ from spirals to BzK galaxies to (U)LIRGs) but would remain sub-linear. A similar correlation between high-J to low-J CO ratio and $<U>$ is found by Wu et al (2014) for resolved SF regions in NGC253.
 It is quite interesting to recall that, following Magdis et al (2012), we can expect 
that $<U>\,\propto SFE/Z$ for our sources. Therefore $<U>$ can be directly related to SFE in this analysis. However, as discussed earlier in this Section, the SFE of local spirals and BzK galaxies is quite similar and cannot justify the different levels of CO excitation. { Hence we conclude that  metallicity ($Z$) differences are probably accounting for an important part of the difference in excitation between local spirals
and $z=1.5$ BzK galaxies. Although we do not have direct metallicity estimates in hand for our targets, we recall that at fixed mass the metallicity of typical SF galaxies is decreasing by over a factor of two from $z=0$ to 2 (see e.g.,
Erb et al 2006; Steidel et al 2014; Maier et al 2014; Zahid et al 2014; Valentino et al 2015)}. However, metallicity likely only plays an indirect role: due to lower gas-do-dust ratios and reduced shielding, it affects the intensity and hardness of the radiation field (and in particular <U>). The stronger radiation field (manifesting as  higher dust temperatures)  then probably directly affect CO excitation. Basically, one can get large $<U>$ values and thus higher CO excitation than in local spirals either  because of the large SFE as in the (U)LIRGs, or also due to the combined role of SFE and Z (as in high-z MS galaxies).

\subsection{Dependences on the SFR surface density}

We provide some further information also on the relation between the CO[5-4]/CO[2-1] ratio and $\Sigma_{\rm SFR}$. The trend
reported in Fig.~\ref{fig:SFE} (right panel, dotted line) scales as:

\beq 
{\rm log}\ I_{\rm CO[5-4]}/I_{\rm CO[2-1]} = 0.26\times {\rm log}(\Sigma_{\rm SFR})+0.27
\label{eq:s_sfr}
\eeq

\noindent
with $\Sigma_{\rm SFR}$ in the usual units of M$_\odot$~yr$^{-1}$~kpc$^{-2}$
Qualitatively the correlation between CO excitation and $\Sigma_{\rm SFR}$ agrees with predictions of the Narayanan \& Krumholz (2014) model, however their calculations for  $\Sigma_{\rm SFR}\sim1$ (as in BzK galaxies)
results in a CO SLED that is somewhat different in shape than that of  the average BzK galaxy. 
In Fig.~\ref{fig:SFE} (right panel, cyan line) we  explicitly report the prediction for the dependence of the CO[5-4]/CO[2-1] ratio on $\Sigma_{\rm SFR}$, which appears to compare nicely to the trend as reported in Eq.~\ref{eq:s_sfr}. 

Therefore, it appears that $\Sigma_{\rm SFR}$ can be used as a good tracer of CO excitation, at least in our sample including local
spirals and (U)LIRGs and high redshift disks. This is not surprising, given the results in the previous sections coupled with the well established correlation between dust temperature (i.e., $<U>$)
and $\Sigma_{\rm SFR}$ (Chanial et al 2007; Hwang et al 2010; Elbaz et al 2011). Explicitly, within our sample the correlation of  $\Sigma_{\rm SFR}$ with  $<U>$ is observed as:

\beq
{\rm log} <U>\ = 0.41\times {\rm log}(\Sigma_{\rm SFR})+1.08
\label{eq:s_sfr}
\eeq

which confirms the idea that 'compactness', equivalent to $\Sigma_{\rm SFR}$, can be also seen as a key parameter driving the physical
properties of galaxies (Elbaz et al 2011; Rujopakarn et al 2011).

\subsection{A linear correlation between L'CO[5-4] and L$_{\rm IR}$ ?}

The availability of CO[5-4] measurements for normal galaxies at $z\sim1.5$ prompts us to re-assess the correlation between CO luminosities
and $L_{\rm IR}$ using such a transition. We recall how the first observations of low-J CO in MS galaxies at $z\sim1$--2 (Daddi et al 2008; 2010; Tacconi et al 2010)
allowed to unveil that such objects have SFRs comparable  to (U)LIRGs but on average three times higher L'CO[1-0]/L$_{\rm IR}$ ratios. The substantially higher CO[5-4]/CO[2-1] ratio in (U)LIRGs suggests that when going to CO[5-4] one might find comparable L'CO[1-0]/L$_{\rm IR}$ ratios  for (U)LIRGs and BzK galaxies.
We show the correlation 
between L'CO[5-4] and $L_{\rm IR}$ in Fig.~\ref{fig:lir54} using the galaxy samples discussed in the previous section, but exploiting individual measurements
here for (U)LIRGs from P12, as opposed to considering only an average value as in Fig.\ref{fig:U}. We also benefit for this analysis from an enlarged sample of 16 additional SMGs taken from the compilation of Carilli \& Walter (2013) and we use the full sample of (U)LIRGs (selected with $L_{\rm IR}>10^{11.3}L_\odot$; 40 objects in addition to the P12 sample) with Herschel/SPIRE/FTS archive data  that will be presented in Liu et al (2015, in preparation).
Quite strikingly, the formal best fitting relation has a slope of $0.96\pm0.04$ and it is thus indistinguishable from the  linear trend:

\beq
{\rm log}\ L'_{\rm CO[5-4]} = {\rm log}\ L_{\rm IR}/L_\odot - 2.52
\label{eq:54}
\eeq

\noindent
(for L'CO in the usual units of K~km~s$^{-1}$~pc$^2$), with a dispersion in the residual of 0.24~dex, which is  remarkable for a relation spanning 4 orders of magnitude along either axes.  We note in particular that for BzK galaxies and P12 (U)LIRGs we find fully consistent luminosity ratios of $-2.36\pm0.05$ and $-2.34\pm0.08$ respectively (we quote the error on the averages, and values are in the units of Eq.~\ref{eq:54}). Both populations have somewhat higher ratios than the average, while spirals have $-2.67\pm0.13$ which is somewhat lower and affected by the strong outlier NGC6946. We are limited though by small number statistics and within the
current uncertainties we cannot detect significant differences among the various populations. A similar conclusion about a seemingly linear slope between L'CO[5-4] and $L_{\rm IR}$ between local (U)LIRGs and high-$z$ SMGs is also presented in Greve et al (2014), although the novelty here is that we can situate the normal MS galaxies at $z=1.5$ in this relation and we include also local spirals (see also Liu et al 2015, in preparation).

The basically linear correlation between L'CO[5-4] and $L_{\rm IR}$ is different from findings for low-J CO, where considering all galaxies together yields a trend more consistent with a slope of 1.7 (e.g., see the classical Solomon \& van der Bout 2005 work) or, when distinguishing SB and MS galaxies,  separate relations with 
slopes of 1.2 and different normalisations (e.g., Sargent et al 2014).  This suggests that CO[5-4] could behave as a tracer of the star forming gas, i.e. the gas that is dense enough to be directly related to the amount of stars formed, in a way that is insensitive on whether the star formation is happening in a quiescent disk or inside a violent merger. If this is the case, then CO[5-4] could be used as a more efficient/accessible tracer of the dense gas than standard molecules like HCN or HCO which also scale basically linearly with  $L_{\rm IR}$ but are intrinsically 10--20 times fainter and harder to observe (Gao \& Solomon 2004ab; Gao et al 2007)\footnote{We note, however, that Garcia-Burillo et al. (2012)  emphasise that (U)LIRGs have a systematically higher $L_{\rm IR}$ to HCN    luminosity ratio  by about a factor of  two  than normal galaxies, and suggest a slightly non-linear correlation between HCN and $L_{\rm IR}$. In our case the comparison of L'CO[5-4] luminosities between BzK galaxies and local (U)LIRGs has the advantage of being done for very similar $L_{\rm IR}$ values.}.
At the same time, given its fairly high frequency, CO[5-4] can be observed with reasonably high, sub-arcsec spatial resolution with both ALMA and IRAM PdBI.  We caution though that the critical density of CO[5-4] is an order of magnitude lower than that of HCN (Carilli \& Walter 2013). Strictly speaking,
 these are not equivalent as tracers.  Also, CO[5-4] has an excitation
 temperature of $\sim$83K, which is much higher than low-J HCN transitions. As such, the
 low-J lines of these molecules are clearer tracers of density,
 because they do not require the medium to be both warm and dense;  we will still need HCN[1-0] or other dense gas tracers at [1-0] to obtain the total dense gas including
the much colder component that will be missed by high J CO lines. Also we do not know if Eq.~\ref{eq:54} manifests a direct connection between
 CO[5-4] and $L_{\rm IR}$ or if both CO[5-4] and $L_{\rm IR}$ are driven in comparable ways by some other parameter, in which case their spatial distribution inside galaxies might differ. Evidence from local galaxies suggest that CO[5-4] emission with Tkin$>50$~K, as favoured by our LVG modelling, might not be excited by PDR but more likely by mechanical heating from shocks (e.g., Rosenberg et al 2014; Meijerink et al 2013), which most likely are driven by winds connected to the SF regions.
Future comparisons of high spatial resolution maps of CO[5-4] with, e.g., HCN, are required to clarify whether this is a viable approach or not,
i.e. to clarify if CO[5-4] can be used as a high fidelity tracer of the star forming gas.
   
\subsection{Implications for gas and dust scaling laws}

In the previous sections we have presented evidences for the existence of correlations, between CO excitation and $<U>$ (eq.~\ref{eq:U}) and between 
$L_{\rm IR}$ and CO[5-4] luminosity, followed by galaxies at $0<z<1.5$, regardless of whether they are MS or starburst  systems (we also include some higher-$z$ SB systems but no $z>2$ MS galaxies). Here we  combine these two empirical trends and examine the
resulting implications. If the relation between CO[5-4]/CO[1-0] and $<U>$ were linear, given Eq.~\ref{eq:54} and that $<U>\propto L_{\rm IR}/M_{\rm dust}$, this would imply L'CO[1-0]/M$_{\rm dust}\sim $~constant. 
However we find that the relation is sub-linear, with a slope of about $\beta=0.7$, which leads to:

\beq
 {\rm log(L'CO[1-0]}/M_{\rm dust})\propto (1-\beta)\times{\rm log}<U>
\label{eq:XU}
\eeq

\noindent 
This relation goes in the right direction of accounting for the 3--5 times higher L'CO[1-0]/M$_{\rm dust}$ ratio observed for (U)LIRGs compared to local starbursts (Magdis et al 2012; Tan et al 2014) given that (U)LIRGs have 10 times higher $<U>$ than spirals, although not fully accounting for the measured difference. Given the observed redshift evolution of $<U>$ 
for MS galaxies (Magdis et al 2012; B\'{e}thermin et al 2014) this relation also predicts a modest rise of the ratio with redshift $\propto (1+z)^{0.4-0.5}$,  consistent with the empirical constraint of Tan et al (2014) who find a trend scaling as $(1+z)^{0.20\pm0.27}$. 

As discussed by Tan et al. (2014), given the fairly similar functional dependence of both $\alpha_{\rm CO}$ and gas to dust mass ratio ($G/D$) on metallicity for local spirals (see also Sargent et al 2014; Bolatto et al 2013; Magdis et al 2012), one would expect a roughly constant $L'CO[1-0]/M_{\rm dust}$ ratio for galaxies if metallicity was the only physical parameter determining
both $\alpha_{\rm CO}$ and $G/D$ in galaxies\footnote{We consider here a fairly high stellar mass regime, $M_{\rm stars}\simgt10^{10}M_\odot$, 
where galaxies are likely H$_2$ dominated.}. This doesn't seem to be the case. If we explicitly take out the metallicity dependence from 
both $\alpha_{\rm CO}$ and $G/D$ assuming they have an identical functional form, $\alpha_{\rm CO} \propto f(Z) \delta_{\rm CO}$ and $G/D \propto f(Z) \delta_{\rm G/D}$, we can re-write Eq.~\ref{eq:XU} in the form:

\beq
 {\rm log}(\delta_{\rm CO} / \delta_{\rm G/D}) \propto (\beta-1)\times {\rm log} <U>
\label{eq:delta}
\eeq

   \begin{figure}
   \centering
   \includegraphics[width=9cm]{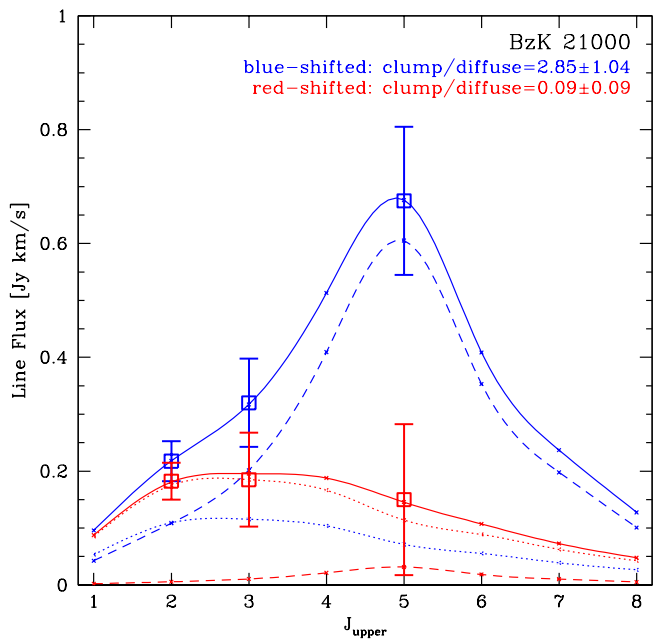}
   \caption{The CO SLED for BzK-21000 is shown separately for its blue- and red-shifted components (using blue and red colours, respectively, points) and decomposed using the 'clumpy' (dashed lines) and 'diffuse' (dotted lines) CO SLEDs from B15 (solid lines are for total SLEDs). The blue-shifted component, corresponding to the northern-half of the galaxy where most massive clumps are located, is more highly excited, requiring only a modest contribution from the diffuse gas. The red-shifted component can be fitted with diffuse gas only.  Given the optical and near-IR morphology of BzK-21000, this is consistent with the predictions of the B15 simulations.
			   }
              \label{fig:diffex}%
   \end{figure}
   
     \begin{figure}
   \centering
   \includegraphics[width=9cm]{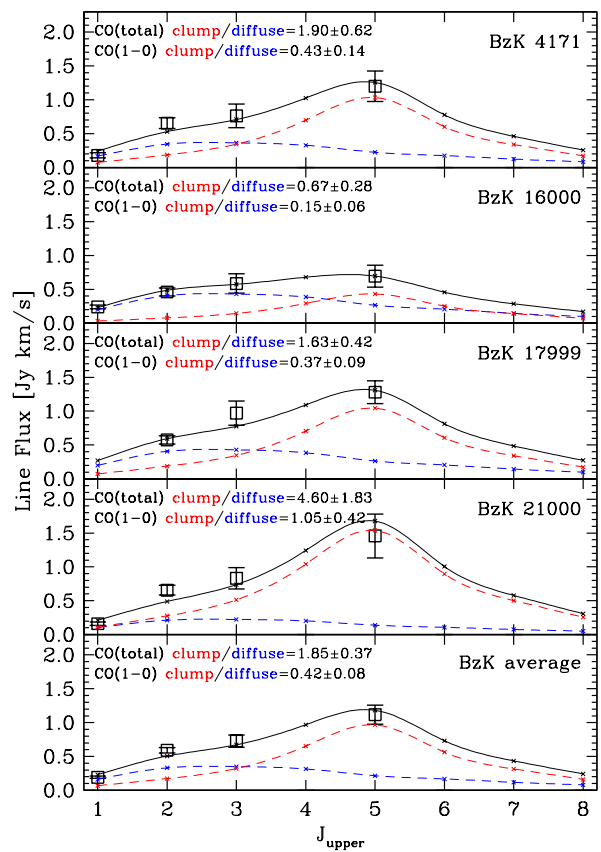}
   \caption{The CO SLEDs of individual BzK galaxies (top four panels) and their average (bottom panel) are fitted using different
   proportions of the 'clump' (red) and 'diffuse' (blue) CO SLEDs from B15. Results of the fit with their uncertainties are reported as labels in the figure, the CO(total) ratio is given relative to the high-redshift disk galaxy simulated in B15. For example, for the average BzK the ratio CO(total) ratio of 1.85 implies that we need a clump component 1.85 stronger than that of B15, if keeping fixed the diffuse component.
			   }
              \label{fig:5pan}%
   \end{figure}
   
\noindent
this relation reasserts again that $z=0$ (U)LIRGs cannot simultaneously follow the same $\alpha_{\rm CO}$ and gas to dust mass ratio ($G/D$) versus metallicity trends as local spirals. The results of Magdis et al (2012) suggest that they mainly deviate low from the $\alpha_{\rm CO}$ versus metallicity relation and follow reasonably well the $G/D(Z)$ trend, but the assessment of metallicity for very dust-embedded 
systems like ULIRGs might be biased low (see Tan et al 2014 for a discussion). Future observations are needed to clarify this issue.

More interestingly, this relation has consequences for the evolution of MS galaxies. Given that their $<U>$ value is rising with redshift,
this implies that at $z>0$ they might not simultaneously follow both the same $\alpha_{\rm CO}$ and gas to dust mass ratio ($G/D$) versus metallicity trend as local spiral galaxies. We cannot distinguish the evolution between the $\alpha_{\rm CO}$  and $G/D$ trend with existing data, as inferring gas masses of distant galaxies relies on adopting at least one of these local trends. However if we were to assume a redshift invariant $G/D(Z)$ relation, eq.~\ref{eq:delta} would imply 0.2~dex lower $\alpha_{\rm CO}$ values at $z=1.5$  than those expected if the local scaling of $\alpha_{\rm CO}$ with metallicity were to apply also to $z=1.5$.

This is similar to the suggestion of Bolatto et al. (2013) that, in addition to the scaling with metallicity, $\alpha_{\rm CO}$ depends as a secondary parameter on $(\Sigma_{\rm Mass})^{\approx -0.5}$ (their Eq.~31; applicable above a density threshold that is practically always matched by the galaxy populations considered here). Because of the MS, this is equivalent to a dependence on ($\Sigma_{\rm SFR})^{\approx-0.6}$ (using a MS slope of 0.8).
Combining our Eq.~\ref{eq:delta} and~\ref{eq:s_sfr} would yield an exponent of $-0.73$, not far from the findings of Bolatto and colleagues. Also, an equivalent scaling with $\Sigma_{\rm SFR}$ is found by Narayanan et al (2012) using simulations. Their Eq.~6 would imply 
an exponent of $-0.4$ (converting from $\Sigma_{gas}$ to $\Sigma_{\rm SFR}$ using a KS relation with a slope of 0.8). 


  \begin{figure}
   \centering
   \includegraphics[width=9cm]{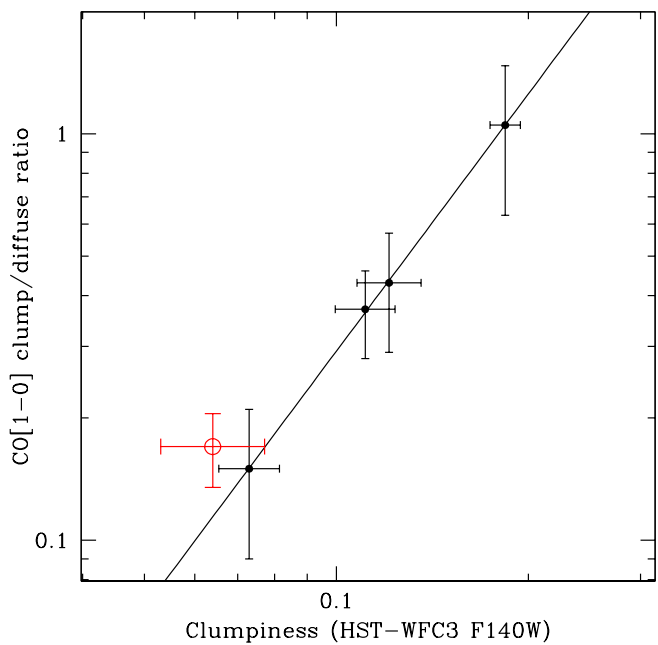}
   \caption{The ratio of CO[1-0] emission decomposed between clumpy and diffuse components (based on CO SLED decomposition, as in Fig.~\ref{fig:5pan}) is shown as a function of the clumpiness of the light as seen
   in the HST-WFC3 F160W imaging (see text for details), corresponding to optical rest frame to $z=1.5$. The red circle shows the properties of the simulated
   disk in Bournaud et al 2014 that appear to fit consistently in the trend (solid line). The B15 model has  a lower clumpiness than the average BzK galaxy.  The fact that the solid-line fit has a $\chi^2\sim0$ (with 2 degrees of freedom) is just a fortuitous property of the sample, with no further parameter adjustment.  
			   }
              \label{fig:clumpiness}%
   \end{figure}

\subsection{Massive star-forming clumps as the driver of the high excitation ?}
\label{sec:clumpiness} 

The galaxy BzK-21000 allows us to investigate the possible role played by the giant clumps in driving the high excitation CO SLEDs of normal
MS galaxies at $z=1.5$, as suggested by B15. Inspecting the CO spectra in the different $J=2,3,5$ transitions in Fig.~\ref{fig:do3} we note that this galaxy is the one with the most obvious variations over the velocity range of the CO emission. The CO[5-4] emission is very weak relatively to CO[2-1] in the red-shifted part, and most of the recovered CO[5-4] signal is in the blue-shifted part. D10
already showed (see their Fig.~6) that the blue- and red-shifted part (relatively to the systemic zero-velocity) corresponds well to the North/South portions of the galaxy, following the  
global rotation in the disk. We averaged  the CO[3-2] and CO[5-4] signal over the red- and blue-velocity ranges separately, as done for CO[2-1],
and extracted CO fluxes for all of them by fitting point-like sources at the  (fixed) positions of the blue and red components of CO[2-1] derived  from D10, which are offset by 0.8$''$. We use point sources here because we are effectively sampling only part of the galaxy each time (so the effective size would be reduced), and also to minimise contamination from the opposite portion of the galaxy. The results of this exercise are shown  in the two panels of Fig.~\ref{fig:diffex}. We measure a  divergence at CO[5-4] (significant at 2.8$\sigma$) and only a hint of a difference at CO[3-2]. 

From Fig.~\ref{fig:wfc3}  (see also Fig.~5 in D10), it is quite obvious that the Northern portion of BzK-21000 is the place where basically all
massive clumps are located, in quantitative support of the analysis of B15. 
We fit the red- and blue- CO SLED of BzK-21000 using the 'clump' and 'diffuse' CO SLEDs derived by B15. We find that the red-shifted emission can be explained using only the 'diffuse' component, while the blue-shifted one is dominated by the 'clumpy' component. The normalization of the diffuse component is similar for the blue- and red-shifted parts, as expected if such a medium is fairly homogeneously distributed within the galaxy.

Encouraged by the possible correlation between clump location and CO SLED for BzK-21000, which supports the idea of B15 that it is giant clumps 
that are driving the high excitation component in high-$z$ MS galaxies, we explored the possibility of accounting for excitation variations among our four 
targets by fitting them with different proportions of  'clump' and 'diffuse' components from B15. Results are shown in Fig.~\ref{fig:5pan}. Our average BzK galaxy requires a larger proportion of 'clump' CO emission, by a factor of 1.8, respect to the simulations 
of B15, likely due to their different gas fractions (see Section~\ref{sleds}). BzK-21000 requires the strongest 'clump' component, and is plausibly the most clumpy galaxy in our sample, at least in terms of massive clumps (Fig.~\ref{fig:wfc3}). Instead BzK-16000 requires 
a smaller proportion, by a factor 0.67, of 'clump' emission than in B15. This galaxy seems the most evolved in our sample, appearing like 
a fairly regular large spiral with a massive bulge and lack of very massive clumps. On average, 30\% of the CO[1-0] emission is coming from the 'clump' warm component, with values for individual galaxies ranging from 13\% for BzK-16000 to 50\% for BzK-21000. 

Figure~\ref{fig:clumpiness} compares the ratio of CO[1-0] between clump and diffuse emission with a direct measurement of clumpiness for our four
BzK galaxies, measured in the CANDELS HST-WFC3 F160W imaging (Fig.~\ref{fig:wfc3}).  
The clumpiness is measured by subtracting from the original images a smoothed version -- obtained through a boxcar averaging with a kernel size equal to 20\% the Kron radius --  and by evaluating the normalized flux in those pixels
 with an intensity larger than 3\,$\times$ the background standard deviation (see, e.g., Conselice 2003; Cibinel et al 2013).
The circular central area (radius=0.2\,$\times$ Petrosian radii) is excluded from the calculation to limit the impact of the unresolved inner galaxy region.    
Errors on the clumpiness are estimated by generating 101 resampled versions of the F160W images in which each pixel flux is replaced with a random value within the photometric error. The clumpiness is re-computed on each resampled image and the dispersion over all realizations provides our uncertainty estimate.
Although based only on four targets, the correlation between clumpiness and clump/diffuse decomposition of the CO SLED is remarkably
good. It should be emphasised that the measurements of clumpiness and CO[1-0] decomposition are totally independent. Good correlations are also found, obviously, between clumpiness and both $<U>$ and CO[5-4]/CO[2-1] ratios.


It is also tantalising to speculate that, if the clumps have much higher CO excitation than the diffuse component in the rest of the galaxy, it might be expected that in future/putative high spatial resolution FIR observations clumps might be found to display higher $<U>$ values than the rest of the galaxies (i.e. warmer FIR SEDs), which might be related to higher SFE  and/or lower metallicities. Both options appear viable at this stage, with lower gas phase metallicity being plausible if they form from collapse of at least partially newly accreted/pristine gas. Evidences for higher SFEs in the clumps  at least in their collapse phases are presented in Zanella et al (2015). Verifying this
possibility will require the combination of multi-wavelength and high resolution tracers of the dust continuum emission, spanning from relatively 
short wavelengths at rest frame $<100\mu$m all the way to the Rayleigh-Jeans tail at 500$\mu$m rest frame or longer. This should be in the
capabilities of ALMA for $z\simgt2$ galaxies, especially if band~10 observations are eventually developed.

 \begin{figure}
   \centering
   \includegraphics[width=9cm]{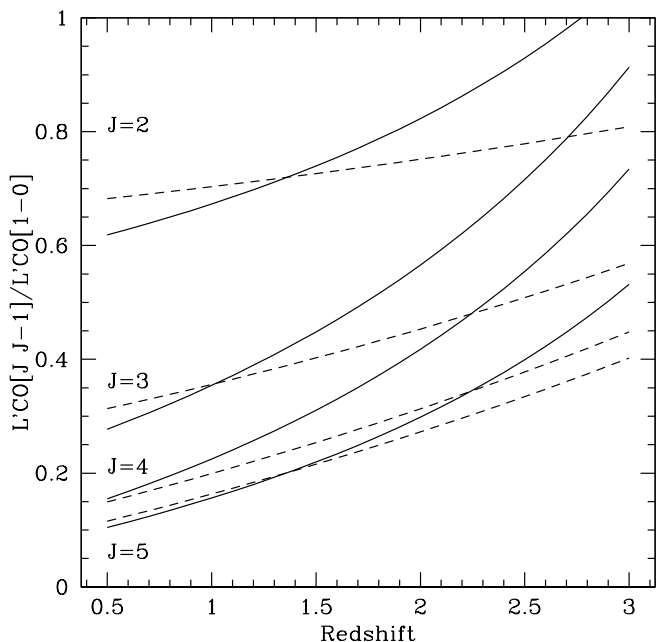}
   \caption{Predicted average luminosity ratios for MS galaxies of various CO transitions normalised to L'CO[1-0], exploiting the correlation between  CO SLED excitation and $<U>$   and between $<U>$ and redshift (Magdis et al 2012; B\'{e}thermin et al 2014). This is computed for the $J=2,3,4,5$ transitions, curves from top to bottom. We considered two component  'cold' and 'warm' CO SLEDs, corresponding to the different solutions shown in Fig.~\ref{fig:lvg}, with the warmer component peaking at $J=5$ (solid) or 7 (dashed).
			   }
   \label{fig:ex_z}%
   \end{figure}

\subsection{Predicting the evolving CO SLED of MS galaxies versus redshift}

The correlation of CO[5-4]/CO[2-1] with $<U>$ can be used to compute expectations for the excitation of normal, MS SF galaxies as a function of redshift. The rise of $U\propto (1+z)^{1.2}$ (Magdis et al 2012)  or possibly steeper (B\'{e}thermin et al 2014) suggests that typical SF galaxies at intermediate redshifts $z\sim0.5$--1 will be less excited on average than those at $z=1.5$, perhaps with CO SLEDs more similar to that of BzK-16000. On the other hand it 
appears that $<U>$ keeps growing steeply all the way to $z\sim4$ at least (B\'{e}thermin et al 2014), suggesting that typical SF galaxies will be more and more excited in their CO emission towards higher redshifts. By $z\sim3$ their intensity of radiation field might get comparable to that of local (U)LIRGs, and similar CO SLEDs might be expected at that point, in large part likely driven by the decline of metallicity towards high-$z$ 
(while we expect this is primarily a SFE effect in (U)LIRGs). 

Following this idea, we have attempted calculations of expected CO SLED versus redshift for MS galaxies. We assume that the SLED can be decomposed
into a 'cold' and a 'warm' component, and vary their relative strength in order to reproduce the trend in Eq.~\ref{eq:U} and Fig.\ref{fig:U}. 
We choose the B\'{e}thermin et al. (2014) functional form of the $<U>(z)$ relation which scales like $(1+z)^{1.8}$. In Fig.\ref{fig:ex_z} we plot 
resulting luminosity ratios, for the case of a warm component peaking at $J\sim5$ (solid lines; Fig.\ref{fig:lvg}) and $J\sim7$ (dashed lines), based on the LVG modelling (Sect.~\ref{lvg}). These tracks can be used to correct CO luminosities of normal MS galaxies in $J\leq5$ transitions to L'CO[1-0]. In applying such corrections to individual galaxies one should consider that, in addition to the uncertainties related to the choice of the underlying 'warm'/'cold' components, the 
dispersion in the $<U>(z)$ relation (of order of 0.2~dex, see Magdis et al 2012) and the intrinsic dispersion in the 
$<U>$ versus CO[5-4]/CO[2-1] trend (0.11~dex in our estimate) imply a scatter of at least 0.15~dex for individual expectations of the CO[5-4]/CO[2-1] ratio. 

\section{Summary and conclusions}

We have obtained CO SLEDs of four normal galaxies at $z=1.5$ including measurements in the CO[1-0], CO[2-1], CO[3-2] and CO[5-4] transitions. All CO[5-4] observations and CO[3-2] observations of 3 galaxies are   presented here for the first time, while the rest of the observations were taken from D10, Dannerbauer et al (2009) and Aravena et al (2010; 2014). This is the first time that such detailed SLEDs are available for normal, un-lensed galaxies, which were selected in the near-IR (not by means of their very high $L_{\rm IR}$) and belong to the star forming main-sequence at their redshift. The measurements are thus representative of typical disk like galaxies at high redshifts, which are the cause for the bulk of the cosmic SFRD.
The main results can be summarized as follows:

\begin{itemize}
\item While the CO SLEDs of all four galaxies are consistent with a Milky-Way-like, low excitation up to CO[3-2] (confirming the results of Dannerbauer et al 2009; Aravena et al. 2010; 2014), we detect substantially excited CO[5-4] emission well in excess of that in local spirals in our sample. The CO SLEDs of $z=1.5$ disks evolve substantially with respect to their local main sequence
counterparts, when considering $J>3$ transitions. However, the CO SLEDs of our sample $z=1.5$ galaxies are less excited than those of local (U)LIRGs and also than the average SMGs, displaying weaker CO[5-4] to low-J CO ratios. 
\item Existing predictions for the SLEDs of this population correctly reproduce the enhancement of CO[5-4] emission with respect to local galaxies.
However, in detail,   P12 predict too highly excited emission,  B15 predict less highly excited emission, while  Narayanan \& Krumholz 2014 predict a shape that is rising too strongly over the CO[3-2]--CO[4-3] transitions. 
Variations observed within individual  galaxies in our sample approach values at either of the extremes of the models.
\item LVG modeling of the SLEDs of the $z=1.5$ galaxies is presented. At least two components are required to reproduce the SLEDs, with the most relevant  constraints applying to their densities. A $n<10^{3}$~cm$^{-3}$ low-excitation component is required, together with a second one at higher density  $n>10^{3.5}$~cm$^{-3}$ with possibly higher temperatures.
\item By comparing different galaxy populations including MS and SB galaxies at low and high redshifts we find that a good tracer of the overall CO excitation, as
parametrized by the CO[5-4]/CO[2-1] ratio, is the intensity of the radiation field $<U>$, which is related to the effective dust temperature. We find a sub linear correlation with $<U>$, having a logarithmic slope of 0.6. The correlation with $<U>$ allows us to predict that intermediate redshift SF galaxies (e.g. at $z\sim0.5$--1) will have intermediate excitation between local spirals and BzK galaxies, while typical $z>3$ SF galaxies might have CO SLEDs with excitations similar to those of local (U)LIRGs. This could counterbalance the lower CO luminosities expected at high redshift due to the higher $\alpha_{\rm CO}$ driven by the declining metallicity, by making higher-J transitions 
more easily observable.
\item We find that the $z=1.5$  normal galaxies follow, within the uncertainties, the same linear correlation between CO[5-4] luminosity and $L_{\rm IR}$ than local spirals, (U)LIRGs and  SMGs. The CO[5-4] emission appears to be related to the star-forming dense gas, regardless of the nature (main sequence vs. starburst) 
of the galaxy. This is  different than what is observed for low-J CO emission. Future observations will clarify whether CO[5-4] can be used as 
a more effective (i.e. less time--consuming) alternative to HCN to study the dense SF gas in galaxies.
\item Within our sample, we find evidence supporting that the high excitation CO emission in the BzK galaxies at $z=1.5$ is originating from their massive clumps, 
in agreement with the predictions of B15.
\end{itemize}

To complete the observations of CO[3-2] and CO[5-4] presented in this paper requested a total on-source time of 82~hours with the IRAM PdBI, while 62~hours were used by D10 to secure high S/N observations of CO[2-1]. Overall, counting standard overheads, this corresponds to a total of about 230~hours of telescope time at PdBI for 3 CO transitions over 4 galaxies. ALMA can now complete such observations more than a hundred times faster, and we hope that the results presented in this paper will be useful to guide the exploration of much larger samples with ALMA. Nevertheless, before ALMA starts observing below 45~GHz, obtaining CO[1-0] flux measurements for $z>1$ galaxies at the VLA will still be a very time consuming endeavour for years to come: over 250~hours were used to secure the high quality CO[1-0] measurements for three galaxies used in this paper. However the results presented here suggest that in most cases CO[2-1] can be a fair substitute, as this transition is on average only marginally sub-thermally excited. This is of course progressively less the case for higher-J transitions, starting from CO[3-2] and especially for CO[5-4], which is seen to be more and more directly sensitive to the SFR rather than to the total H$_2$ gas reservoir.


%

%

\begin{acknowledgements}
We thank Ronin Wu and Aur\'elie Remy-Ruyer for providing information for M83 and IC342. We are grateful to Desika Narayanan and Padelis Papadopoulos for discussions and to the anonymous referee for a constructive report.
This work was based on observations carried out with the IRAM PdBI, supported by INSU/CNRS(France), MPG(Germany), and IGN(Spain). The National Radio
Astronomy Observatory is a facility of the National Science Foundation
operated under cooperative agreement by Associated Universities, Inc. 
We acknowledge the use of GILDAS software (http://www.iram.fr/IRAMFR/GILDAS). We acknowledge the support of the ERC-StG UPGAL 240039 and ANR-08-JCJC-0008 grants. 
\end{acknowledgements}



  


\end{document}